# Lessons learned from Milan electric power distribution networks data analysis during COVID-19 pandemic

Alessandro Bosisio
*Energy Department*
*Politecnico di Milano*
Milan, Italy
alessandro.bosisio@polimi.it

Francesca Soldan
*Transmission and Distribution Technologies Department*
*RSE SPA*
Milan, Italy
francesca.soldan@rse-web.it

Andrea Morotti
*Planning Department*
*Unareti S.p.A.*
*Milan, Italy*
andrea.morotti@unareti.it

Samuele Grillo
*DEIB Department*
*Politecnico di Milano*
Milan, Italy
samuele.grillo@polimi.it

Enea Bionda
*Transmission and Distribution Technologies Department*
*RSE SPA*
Milan, Italy
enea.bionda@rse-web.it

Gaetano Iannarelli
Department of Astronautics, Electrical and Energy Engineering
*Sapienza University,*
*Rome, Italy*
iannarelli.800040@studenti.uniroma1.it

***Abstract*—COVID-19 pandemic has been a disruptive event from health, social, and economic points of view. Besides that, changes in people's lifestyles, especially during the 2020 lockdowns, also affected energy networks. COVID-19 pandemic has resulted in a significant decline in electricity demand. The lockdown measures applied to handle the health crisis have caused the most relevant energy impact of the last years. In this paper, the local experiences of the distribution network of Milano, a city in northern Italy, are reported. The analysis starts with a summary of the restrictions imposed during 2020 and focuses on both active and reactive power flows, and faults. To this end, a comparison with 2019 data has been performed, highlighting the main differences with 2020. The outcome of the analysis is a valuable tool to predict urban distribution networks behavior during times of disruption, helping distribution system operators to prepare feasible short-term and long-term resilience plans.**

***Index Terms*—COVID-19 pandemic, distribution networks, power distribution faults, reactive power, resilience.**

## I. Introduction

Coronavirus SARS-CoV-2 emerged in Wuhan, a city in China's Hubei province, in late 2019. Coronavirus disease 19 (COVID-19) rapidly spread with confirmed cases in almost every country worldwide and has become a new global public health crisis. In Italy, the northern regions (Lombardia, Veneto, and Emilia-Romagna) have been the first areas most affected by the epidemic, with thousands of infected people and deaths. To slow down the infection, the Italian government started a national lockdown on March 9$^{th}$, 2020, with enormous social and economic impacts on the entire country. During the lockdown, the government imposed increasing preventive measures, which involved closing schools and churches, shops, bars, restaurants, non-essential firms, and industries.

Some studies on the effects of the COVID-19 pandemic on power systems have already been published. In [1], the authors review recent literature related to the effects of the COVID-19 pandemic on energy systems and electric power grids. The paper points out the main challenges that the pandemic introduced by presenting electricity generation and demand patterns, frequency deviations, and load forecasting. Moreover, the authors suggest directions for future research that may assist in coping with the mentioned challenges. Data analyzed in [2] show that COVID-19 strongly affects global energy systems. Global power sector $CO_2$ emissions have shown a substantial decline due to the COVID-19-induced economic downturn and resulting reduction of electricity demand. Authors in [3] show that the nation-and region-wise lockdown imposed to reduce the COVID-19 has affected businesses, industries, and transportation, resulting in a change in the electric load demand pattern. Due to the changes in work patterns and lifestyles, residential electricity demand has increased. In contrast, industrial and commercial load demand has reduced. The study investigated various issues and challenges faced by the utilities in the Indian power system, including the essential measures taken to retain the grid frequency and voltage profile within their recommended bands. In [4], high-resolution, disaggregated data are analyzed to measure the shifts in electricity use related to Heating, ventilation, and air conditioning (HVAC) loads, non-HVAC loads, and whole-home loads in a comparison of 225 housing units over the years 2018–2020. Key findings from the analyses indicated increased electricity use during periods that occupants would usually be away from home. The most percent increases in non-HVAC residential loads occurred between 10 a.m. and 5 p.m.; HVAC loads increased in total daily consumption compared to the same average daily temperatures of previous years. Additionally, by dividing the data by household income, both lower- and higher-income households experienced the more significant increases in consumption, while the middle-income groups experienced smaller-scale shifts.

Considering east and middle-east countries, in [5], the impact of a reduction in load demand on renewable energy in the Japanese public power grid under a state of emergency declaration, April to May 2020, is studied. COVID-19 caused reductions in load demand of different magnitudes in 2020. The closure of large factories and the limitation of people's activity due to the emergency declaration resulted in a considerable reduction in working time. Moreover, electricity companies took countermeasures to respond to the load reduction due to the pandemic. These measures include adjusting the traditional generation of thermal nuclear power plants, the penetration of renewable energy power, and the utilization ratio of dispatch sectors, such as pumped hydro storage systems and transmission lines between neighborhood regions. Photovoltaics (PV) penetration increased with the load demand reduction in the research areas. The overall load decline did not cause a decrease in PV power generation, and its penetration rate increased, improving the utilization rate of PV. Finally, electricity spot trading prices decreased with the increased proportion of PV power generation. PV power generation reshaped the distribution pattern of the grid load, impacting the electricity trading price. In [6], the authors investigated the impact of the pandemic on the spatial patterns of electricity consumption in six socioeconomic sectors, i.e., residential, industrial, commercial, government, and productive farms, in the State of Qatar. The authors assessed the spatiotemporal patterns of electricity consumption using various Geographic Information Systems (GIS) and spatial statistical modeling before and during the pandemic. The findings of this study showed that the industrial and commercial sectors were the most affected by the pandemic. There was a substantial reduction in electricity consumption in these sectors, indicating that the country's economy has been negatively affected by the pandemic and the associated measures taken to slow down the spread of the infection. Conversely, the production farms sector was not affected by the spread of the disease. The electricity consumption in the residential and the government sectors witnessed a substantial increase during the summer months only during the pandemic year. To better detect the impact of the COVID-19 pandemic on energy consumption, paper [7] compared pandemic-free scenarios with actual (with COVID-19) energy consumption in 2020, rather than comparing energy consumption between 2020 and 2019 in the existing studies. The approach used for scenario simulation combined autoregressive integrated moving average (ARIMA) models and backpropagation neural networks (BPNNs). The proposed simulation approach was run based on China’s electricity consumption from 2015 to 2019 to produce the simulated value of China’s electricity consumption from January to August 2020 in the pandemic-free scenario. On average, the actual electricity consumption was 29% lower than the electricity consumption in the pandemic-free scenario, which is greater than the decline rate derived from the year-to-year comparison. In addition to this, the results of the correlation analysis show the simulated decline in electricity consumption is only positively correlated with the number of new cases of COVID-19 in January–March when the COVID-19 outbreak in China.

At the European level, the paper [8] presents a systemic approach toward assessing the impacts of the COVID-19 pandemic on the power sector. The analysis shows that the most immediate consequences regarded the power demand profiles, the generation mix composition, and the electricity price trends. The decrease in electricity demand resulted in a reduction of the conventional dispatched power. This led to an increment of renewable energy source (RES) generation over the total one. An increase of the non-conventional generation amplifies the operational challenges and the need for regulation capabilities to keep frequency stability and procure resources for the voltage regulation. The impact on the electrical market is twofold. On one side, the load reduction eliminated the need for the most expensive conventional power to balance the demand in the day-ahead market, resulting in lower electricity market prices. On the other side, transmission system operators had to promptly dispatch conventional power generation units, whose services are acquired from the ancillary service market, increasing the cost for ancillary services. In [9], the authors assess the impact of different containment measures taken by European countries in response to COVID-19 on their electricity consumption profiles. The comparisons involved Spain, Italy, Belgium, and the UK (countries with severe restrictions) and the Netherlands and Sweden (countries with less restrictive measures). The results show that different lockdown measures in European countries and their effects on population activities have considerably changed the consumption profiles. For countries with severe restrictions, such as Spain, Italy, Belgium, and the UK, weekday consumption decreased and energy consumption profiles are similar to pre-pandemic weekend profiles for the same period in 2019. However, the reduction in power consumption was lower for countries with less restrictive measures. As a matter of fact, for Sweden, which did not impose a lockdown, the consumption, if compared to the same period in 2019, increased at specific moments. These outcomes imply that while lockdown measures change the consumption profiles on the one hand, on the other hand, these profile changes can also be a reflection of the effects of different approaches to dealing with the pandemic on people’s activities. Authors in [10] described the changes in the power system of the COVID-19 pandemic using the case study of Great Britain. The changes were characterized with various quantitative markers and compared with pre-lockdown business-as-usual data. The outbreak of COVID-19 disrupted the patterns in electricity consumption, challenging the system operations of forecasting and balancing the supply and demand. The ripple effects on the generation portfolio revealed a 5 to 10% higher RES share and a decreased usage of conventional plants. The system stability indicators suggest that the grid operated well but was under stress. The indicators used in the paper include some remarkable loss of load probability events and overall higher system frequency. However, other contrasting findings show 3 to 5% more accurate day-ahead load forecasts. The energy market is also greatly affected by the change in consumption patterns. The wholesale market price decreased due to RES generators ranking higher on the merit order. While the imbalance prices increased due to the higher imbalance volume in the system, this increase was compensated by the larger number of available generators due to the decreased demand volume. Paper [11] focuses on the effects of the first wave of the COVID-19 pandemic on the electricity sector, in particular in Germany. The COVID-19 pandemic demonstrated that it is possible to integrate an increased share of RESs into a changing electricity system at a continuously high level of security of supply. From a market perspective, the paper shows that the decreased electricity consumption and a higher share of RES led to a lower price level on the day-ahead market compared to previous years and an increased number of hours with negative electricity prices. In [12], the authors study the consequences of the economic shutdown on the Iberian electricity market and discuss the timeline of events, the macroeconomic outlook, the financial status of the major electric utility companies (before being hit by the COVID-19 health pandemic), the changes in load profile, the generation mix and, finally, the electricity market spot prices.

At the Italian level, the study's goal presented in [8] is to show the impact on the power industry of all the restrictions and lockdown of the activities in Italy and discuss the effects of the COVID-19 outbreak on the bulk power system and the entire electricity sector. In particular, the consequences on load profiles, electricity consumption, and market prices in Italy, including the environmental aspects, are examined in the paper. The study shows that the pandemic caused a reduction of consumption of 37% compared to the same time of the previous year. The reduction in consumption excluded part of the costly thermoelectric generation, which is automatically taken out of the market due to the production cost compared with renewable generation. Consequently, wholesale energy prices decreased about 30% in the last weeks of March and in the first week of April, and in some cases reached a value of 0 €/MWh. This circumstance had immediate consequences also on the reduction of $CO_2$ emissions. Authors in [13] investigate the impact of the lockdown on load and production in the distribution system of Terni. The data analysis reveals that, even if the emergency related to the COVID-19 pandemic brings no relevant issues for the distribution network management, the lockdown globally brings a consumption reduction compared to the previous years. The consumption reduction upsets the typical demand pattern in domestic customers' behavior. Referring to primary substations power flows, the transmission system operator could find unexpectedly intense reverse power flows, with a distributed generator's production surplus and voltage profile issues. Paper [14] analyzes the impact of the COVID-19 pandemic situation on residential loads and local distribution transformers. The authors observed an increase in energy consumption during the entire office hour or part of the office hour. As a result, the residential transformer gets vulnerable since the value of the power increases. Moreover, the higher harmonic loss factor tightens the range of power values for the transformer to operate safely. Lastly, paper [15] treats the critical topic of energy insecurity. The COVID-19 pandemic and associated changes in social and economic conditions affected the prevalence of energy insecurity. The authors stress the essentiality of providing relief to the growing number of energy-insecure households and protecting them from even more dire circumstances caused by utility disconnections and unpaid energy bills.

Based on the presented literature review, the main effects of the COVID-19 pandemic on the electrical power systems were a change in electricity demand in terms of intensity and patterns, a change in the power generation mix, in $CO_2$ emissions, and in electricity market prices. In this paper, the local experiences of the distribution network of Milano, a city in northern Italy, are reported. The analysis mainly focuses on active and reactive powers and faults. For doing so, a comparison with pre-lockdown business-as-usual data has been performed, highlighting the main differences with 2020. The paper is organized as follows: Section II briefly reports the 2020 timeline and related countermeasures. In Section III, an overview of the data analyzed is described. Section IV shows the outcomes of analyzing Milano’s active and reactive power data, while Section V shows up the results regarding network faults. Concluding remarks are given in Section VI.

## II. COVID-19 LOCKDOWN: AN OVERVIEW OF THE POLICY RESPONSES

In the following discussion, we investigate possible COVID-19 impacts on the Unareti Milan distribution grid. The analysis is done in view of the main ministerial decrees and regional ordinances of 2020, which imposed mobility, activity, and social restrictions and therefore brought multitude disruptions to daily life. For the sake of clarity, we have divided the whole 2020 year into seven steps:

- Step 1 (1/02 - 22/02): no COVID-19 cases are still found in Italy.
- Step 2 (23/02 - 9/03): after the first COVID-19 cases and local closures, the ministerial decree of March 1st 2021 [16] (“*Additional provisions implementing Decree Law No. 6 of February 23, 2020 on urgent measures concerning the containment and management of the epidemiological emergency by COVID-19*”) imposes restrictions in some regions of Northern Italy (Lombardy, Veneto, and Emilia Romagna) and a limited number of provinces in central Italy. The city of Milan is therefore included in this first closure.
- Step 3 (10/03 - 3/05): the Ministerial Decree of March 9th [17] establishes the first lockdown in the entire Italian country, starting from March 11th. It is allowed to go out only for working or health reasons, and to go buying essential goods. From March 22nd, all non-essential activities are limited.
- Step 4 (4/05 - 6/07): some commercial activities open again, but social distance prescriptions and the ban of gatherings are still in force. Between May and June, most activities can start again: restaurants, cafes, cinemas, music halls, sports centers, etc…
- Step 5 (7/07 - 12/10): after a summer without restrictions, COVID-19 cases start their increase again in September. However, schools open with new rules to limit infections.

- Step 6 (13/10 - 2/11): new Ministerial Decrees [18] try to stem the second COVID-19 tide of infections. The closures are gradual and start from conventions and fair trades; finally, cultural and activities are limited again by additional Ministerial Decrees [19], [20], [21].
- Step 7 (3/11 - 31/12): curfew is established in the entire country from 10 pm to 5 am. During the Christmas holidays, new lockdowns are established.

It is worth noticing that the urban area of Milan mainly has a residential/commercial nature. The customers can be clustered as follows: 905000 LV customers (with total contractual power of 5400 MW), 1700 MV customers (with total contractual power of 1625 MW), ten heavy-industrialized customers, which count for 60MW, only 0.85% of the total contractual power. Therefore, even if the restrictions were imposed on most activities, since heavy-industrialized customers only represent a small part of the power, we can reasonably argue that the changed operation of the distribution network during the COVID-19 pandemic is not strictly related to the change in behavior of this type of loads.

## III. INPUT DATA OF THE ANALYSIS

The analysis reported in the paper is based on electrical measurements and information about outages on the Unareti Milan distribution network, combined with topological, geographical, and meteorological data. The integration of different data sources provided a comprehensive view of the Milan grid between 2019 and 2020. 2019 is considered a reference period to understand which effects the COVID-19 pandemic brought on the Milan distribution grid in 2020. In the following, a description of available data is given.

### *A. Electrical measures*

Electrical measurements refer to real and reactive power sampled with a time resolution of 15 minutes at the 37 High Voltage/Medium Voltage (HV/MV) transformers of the Unareti network. Data availability is high, as most transformers have less than 5% missing data in 2019 and 2020. Table II reported in Appendix A shows the percentage of missing data for each transformer separately for real power and reactive power. Only a few transformers are critical. However, when comparisons between the two years are necessary, we have selected only those timestamps for which data in 2019 and 2020 are available.

### *B. Outages data*

Data are available from 2012 to 2020 in a format like that reported in Table I. It is worth noticing that, since most of the faults affected the MV section [22], only outages related to the MV networks have been taken into account. For each outage, the primary pieces of information stored are: topological data, primary substation, HV/MV transformer, MV busbar, and MV feeder supplying the outage component; temporal data, outage starting and ending timestamps; classification data, which could be either of identified or unidentified origin; location data, the network section and the damaged component. Having the upstream HV/MV transformer for each damaged component makes it possible to relate outages to the electrical measures. Table II shows the number of outages referred to each HV/MV transformer reported for the years 2019 and 2020.

TABLE I. FAULTS DATABASE

| **Primary substation** | **HV/MV transformer** | **MV busbar** | **MV feeder** | **Fault date start** | **Fault date end** | **Classification** | **Power system section** | **Power system component** |
|---|---|---|---|---|---|---|---|---|
| PS1 | HV/MV1 | Busbar1 | Feeder1 | 02/01/2020 08:51:33 | 02/01/2020 09:21:48 | Identified origin | MV primary network | HV/MV transformer breaker |
| … | … | … | … | … | … | Unidentified origin | MV secondary network at 23kV | HV/MV transformer |
| … | … | … | … | … | … | … | MV secondary network at 15/9/6.4kV | MV feeders breaker |
| … | … | … | … | … | … | … | … | 15/9/6.4kV cable |
| … | … | … | … | … | … | … | … | 23kV cable |
| … | … | … | … | … | … | … | … | 23kV secondary substation transformer |
| … | … | … | … | … | … | … | … | 15/9/6.4kV secondary substation transformer |
| … | … | … | … | … | … | … | … | MV electrical cabinet |
| … | … | … | … | … | … | … | … | MV/MV transformer |
| … | … | … | … | … | … | … | … | Interconnection cables breaker |
| … | … | … | … | … | … | … | … | Interconnection cables |

### C. *Geographical and topological data*

The electric network of Milan is represented using the standard IEC CIM [23], [24]. Topological, asset, geographical, and diagram information are saved in a semantic data store, queryable using the SPARQL language. Fig. 1 represents the secondary substations of the Milan Unareti network. The different colors indicate the nominal voltage of the belonging feeders; the circle sizes are proportional to the number of outages in each feeder in 2019 and 2020.

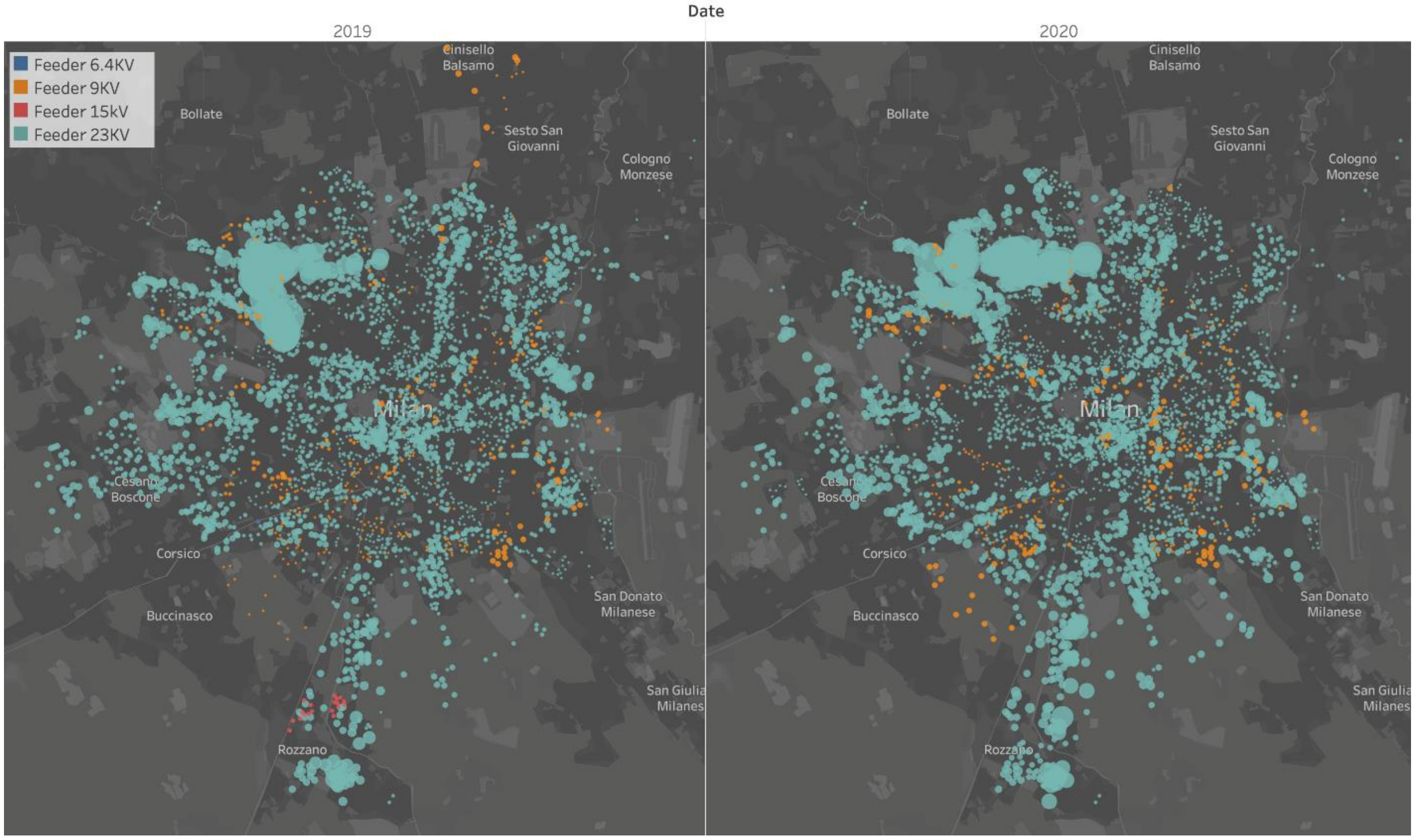

Fig. 1 Map of the secondary substations of Milan Unareti network. The feeder nominal voltage is indicated using a different colour. At the same time, the size of the circles representing the secondary substations is proportional to the total number of outages that occurred in the corresponding feeders.

### D. *Meteorological data*

For the scope of our work, we have limited the analysis to temperature and rainfall measured at the meteorological station of Milan - Juvara. This station is located in the city center of Milan and belongs to the meteorological network of the Regional Environmental Protection Agency (ARPA Lombardia) [25]. Behind the choice of these meteorological data is the hypothesis of their more substantial influence on the behavior of electrical variables and outage occurrences [22]. Fig. 2 represents the monthly trend boxplot of meteorological variables for the two years. The boxplots show the maximum and minimum of the data set, the median, the first, and the third quartile. Moreover, the monthly average values are also reported using green and red lines. It is evident a very steep increase of temperature from May to June in 2019, while in 2020, this temperature rise is more gradual. In the days between 26 and 29 June 2019, an especially notable heatwave has been recorded, with a minimum-maximum temperature range within 27.6 and 38.9 °C. For what concerns rainfalls, the highest value in 2019 was recorded in August, while in 2020, the summer months were rainier, and the highest value was recorded in July.

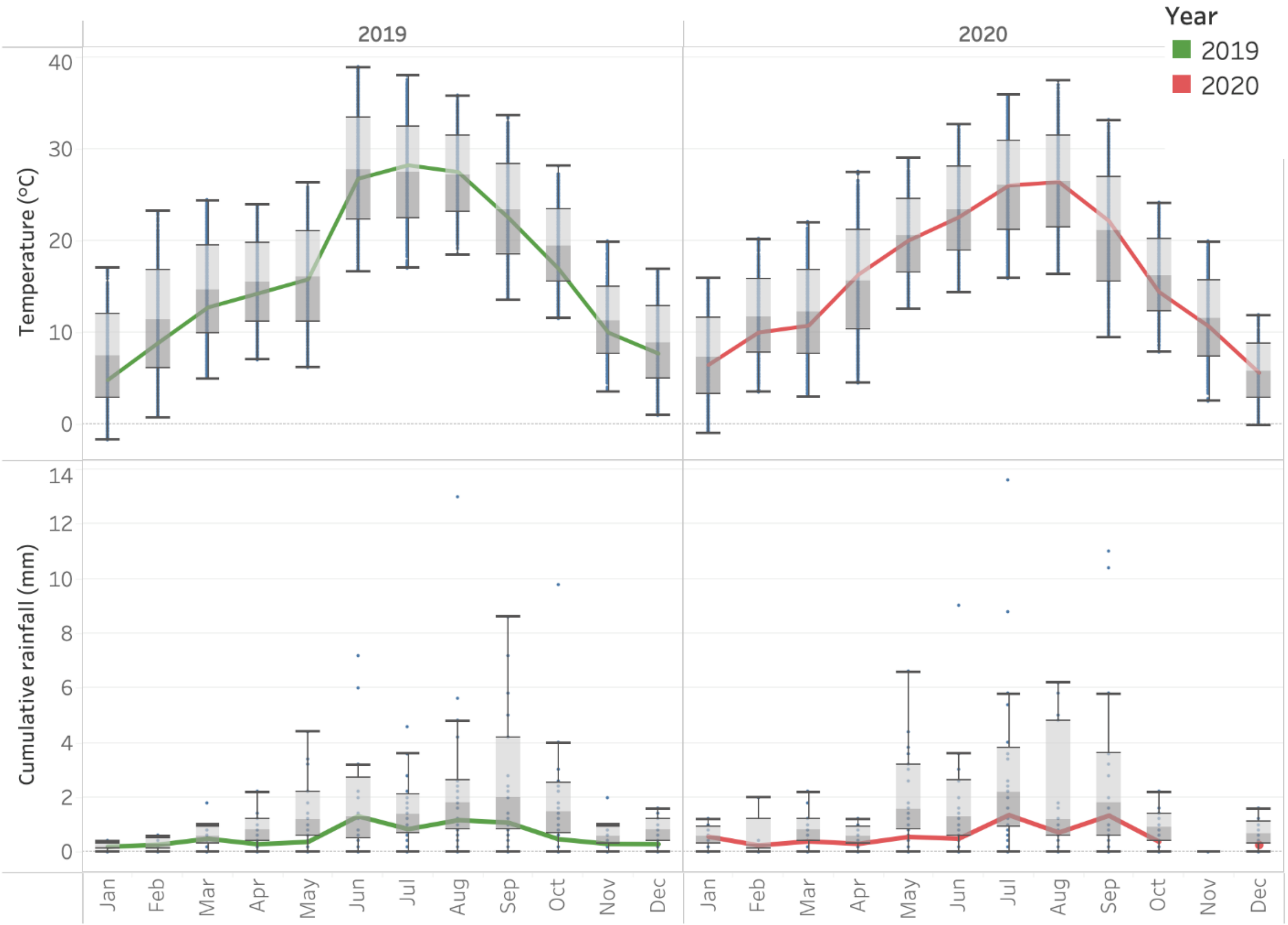

Fig. 2 Boxplots of temperature and cumulative rainfall variables for 2019 and 2020. Data are collected at the ARPA meteorological station of Milan – Juvara with a time resolution of 10 minutes. The solid line indicates monthly average values (excluding 0 mm measurements for rainfall).

## IV. COVID-19 LOCKDOWN IMPACT ON ELECTRICITY DEMAND: ACTIVE AND REACTIVE POWER

This section reports the primary evidence returned from the analysis of active and reactive power trends. The data of 2019 and 2020 are reported and compared considering the seven steps presented in Section II. Even if in the input database active and reactive power records are available for each HV/MV transformer, for the sake of simplicity, the charts report only the average values computed over the data of all transformers.

### *A. Active Power*

The 2021 Unareti Development Plan [22] compares the Milan grid's electrical consumptions between 1/01/2019 and 31/05/2021. The main observation is a significant decrease in demand in 2020, around –8.6%, and a further increase in 2021, +23.9%, compared to 2020. However, the power demand does not reach the value registered in 2019. Still, it is –1.9% compared to 2019. This behavior depends on the city's commercial load profile and the diffuse resort to remote working and remote university lectures. Fig. 3 shows the average HV/MV transformer's real power trend in 2019 and 2020. Seven vertical lines are also drawn to mark starting and ending dates of the steps defined in Section II. Furthermore, the monthly average active power is shown to highlight the changes recorded in 2020 compared to 2019.

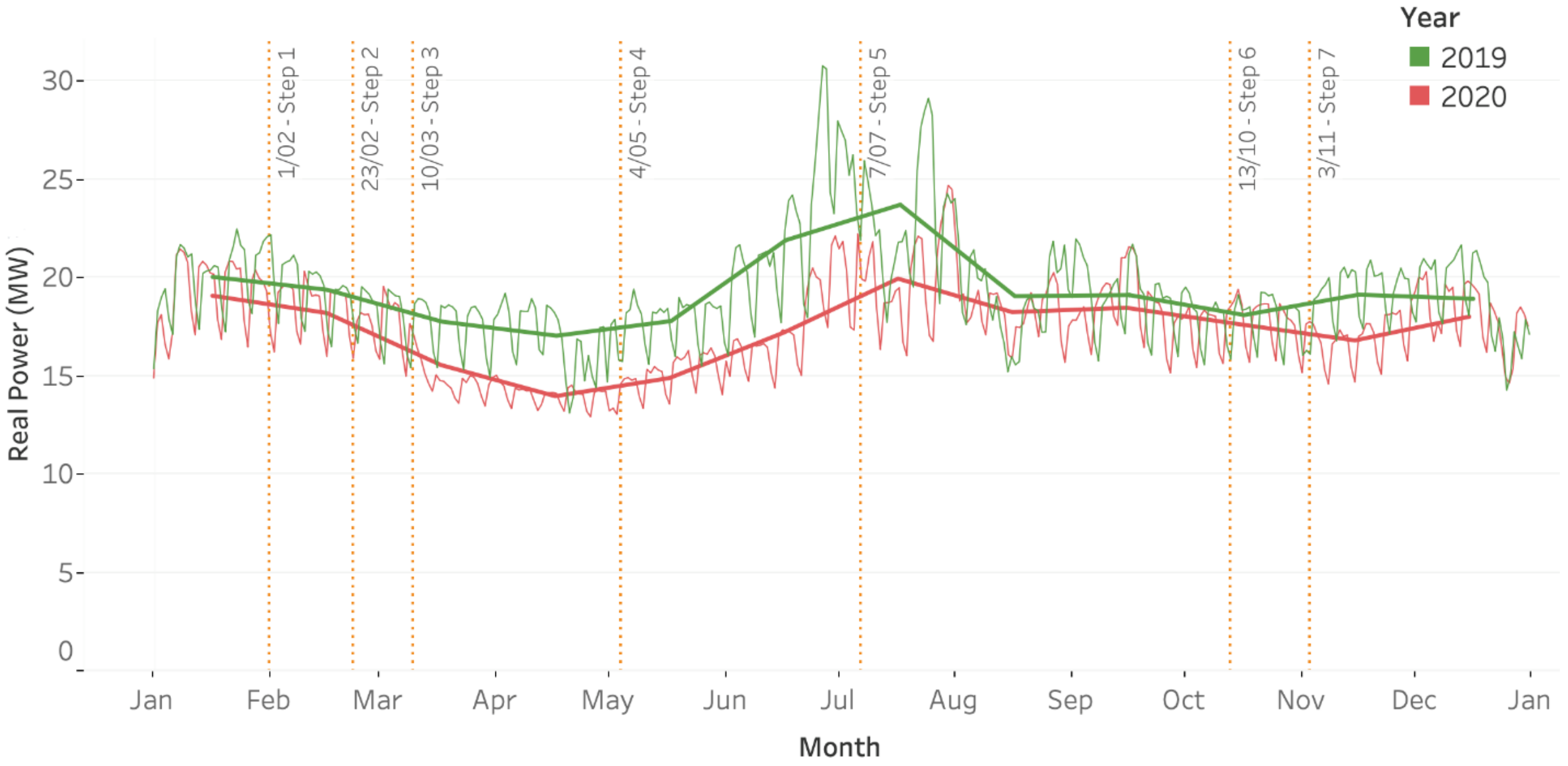

Fig. 3 Average HV/MV transformer's real power in 2019 and 2020.

On the one hand, the 2019 curve has the traditional layout of five business days and two lower demanded weekend days. Moreover, a seasonal trend clearly shows up: in wintertime, the power demand is slightly higher than in spring and autumn; due to the usage of air conditioners, the demand starting to increase from May, reaching the peak in July; a valley in the middle of August identified the usual summer vacations. On the other hand, the 2020 load profile has been affected by the consequence of the COVID-19 pandemic. The comparison between the two years is better outlined in Fig. 4. It emerges that:

- Even in the first months of 2020, it is evident a slight reduction of active power, which is however lower than 5% (Step 1 (1/02 - 22/02) and Step 2 (23/02 - 9/03). The reduction is mainly related to two factors: (i) the different calendar of working and weekend days; (ii) as shown in Fig. 2, February 2020 ranked globally as the second warmest month since 1880. Therefore, the higher temperatures recorded in winter 2020 make the demand lower than in 2019;
- A higher reduction of active power starts from the close of businesses at the beginning of March and reaches values of 18% in Step 3 (10/03 - 3/05);
- Power demand reduction continues in Step 4 (4/05 - 6/07), where the variation between the two years is around 21%. From May, the demand increases in both years and reaches its maximum values in the summer months due to the highest consumptions linked to the use of air conditioning systems;
- In Step 5 (7/07 - 12/10), the power demand in 2020 is, on average, lower than 2019 by 5%. The reduction is concentrated in July, while September and August do not show any remarkable variation;
- In the remaining months, Step 6 (13/10 - 2/11) and Step 7 (3/11 - 31/12), the variation between the two years becomes lower, but consumptions in 2020 never reach the one of 2019.

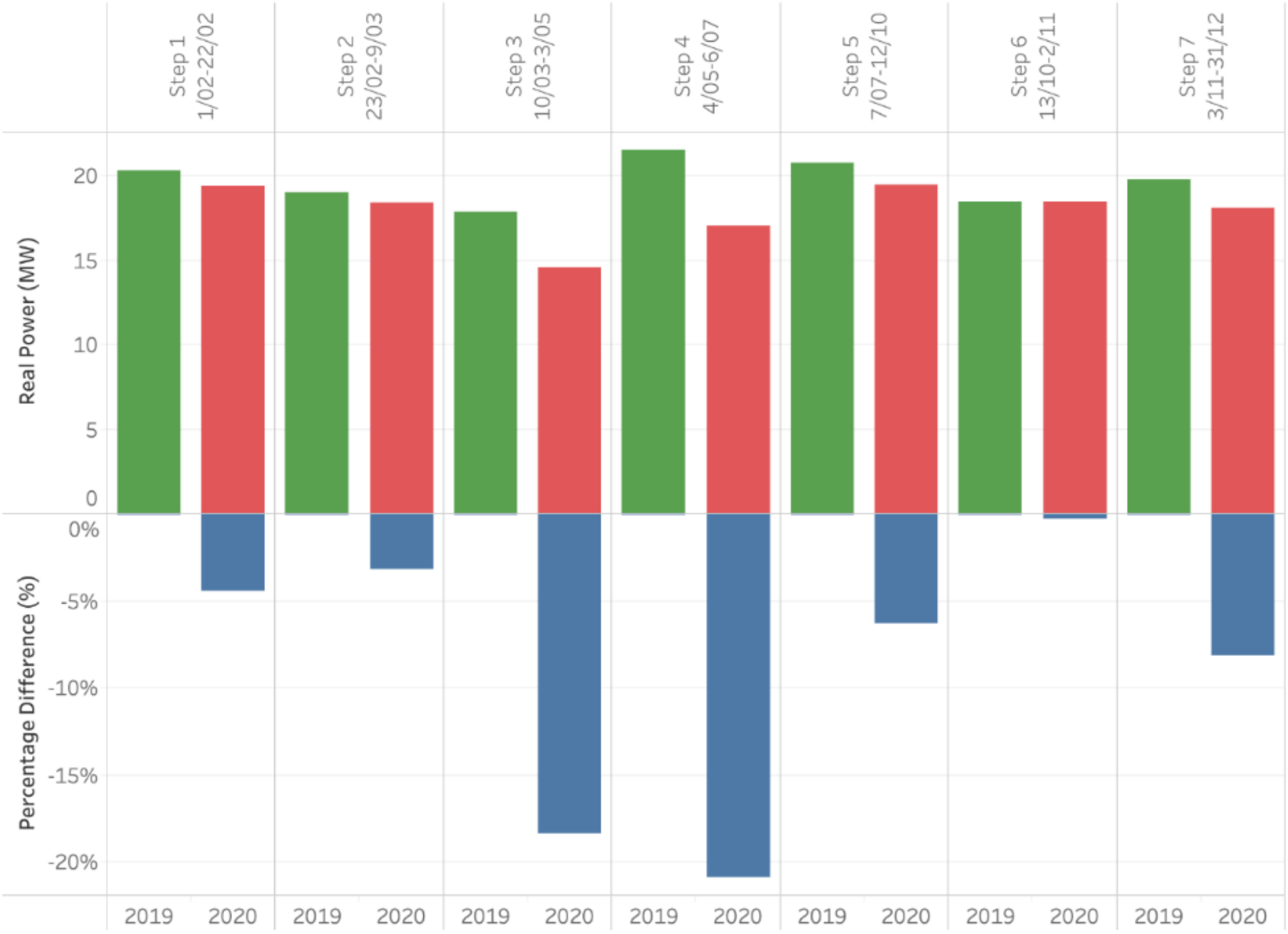

Fig. 4 Average real power measured in each of the seven steps. 2019: green; 2020: red. In blue is the variation.

Papers [26] and [27] have already investigated changes not only in terms of demand intensity but also of load profiles on the Unareti grid. Authors considered consumptions on specific MV feeders respectively for the case of Milan and Brescia, for the first half of 2020. In this paper, we extend this kind of analysis to the entire year 2020, focusing on aggregated data and not on single MV feeders. On the one hand, the main observations for a typical business day are shown in Fig. 5:

- In the first two steps (Step 1 (1/02 - 22/02) and Step 2 (23/02 - 9/03), the average demand profile of 2020 has the exact shape of the demand profile of 2019, with the highest peak around 10 am, followed by a valley in the central hours of the day and a further increase in evening hours;
- Step 3 (10/03 - 3/05) shows a reverse pattern instead, with the highest peak in the evening. This profile has the shape of a typical residential user, confirming the closure of most industrial/commercial activities and the more prominent presence of people at home;
- In Step 4 (4/05 - 6/07) and Step 5 (7/07 - 12/10), air conditioners affect the trend. The two-peaks behavior of previous months disappeared, due to the use of air conditioning also in the central hours of the day;
- In Step 6 (13/10 - 2/11) and Step 7 (3/11 - 31/12), the daily pattern comes back to business two-peaks typical profile.

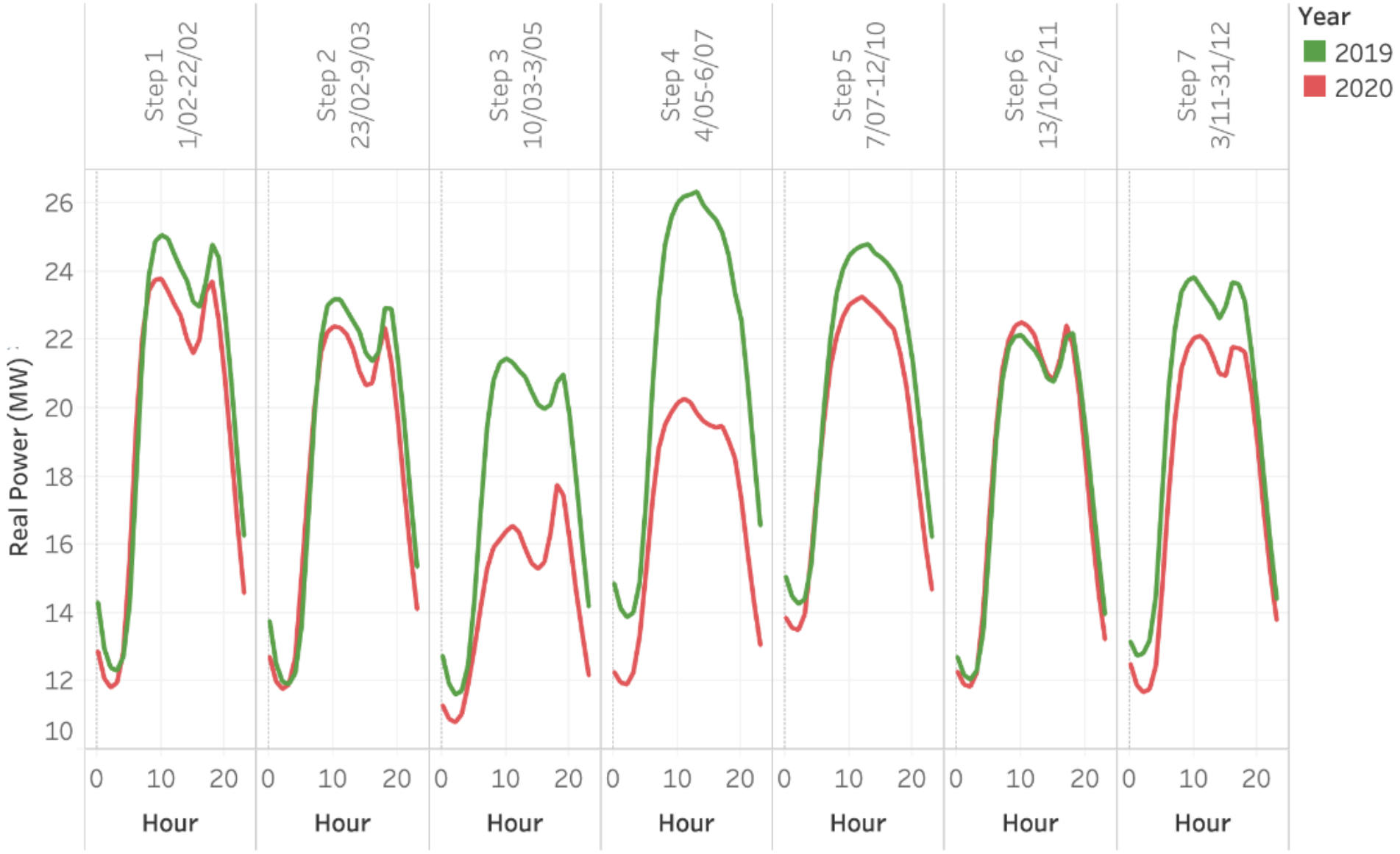

Fig. 5 Business days real power profiles in the seven steps for 2019 (in green) and 2020 (red line).

On the other hand, it is interesting noting that, during weekends, the average demand profile keeps the same shape, with the typical two daily peaks shown in Fig. 6.

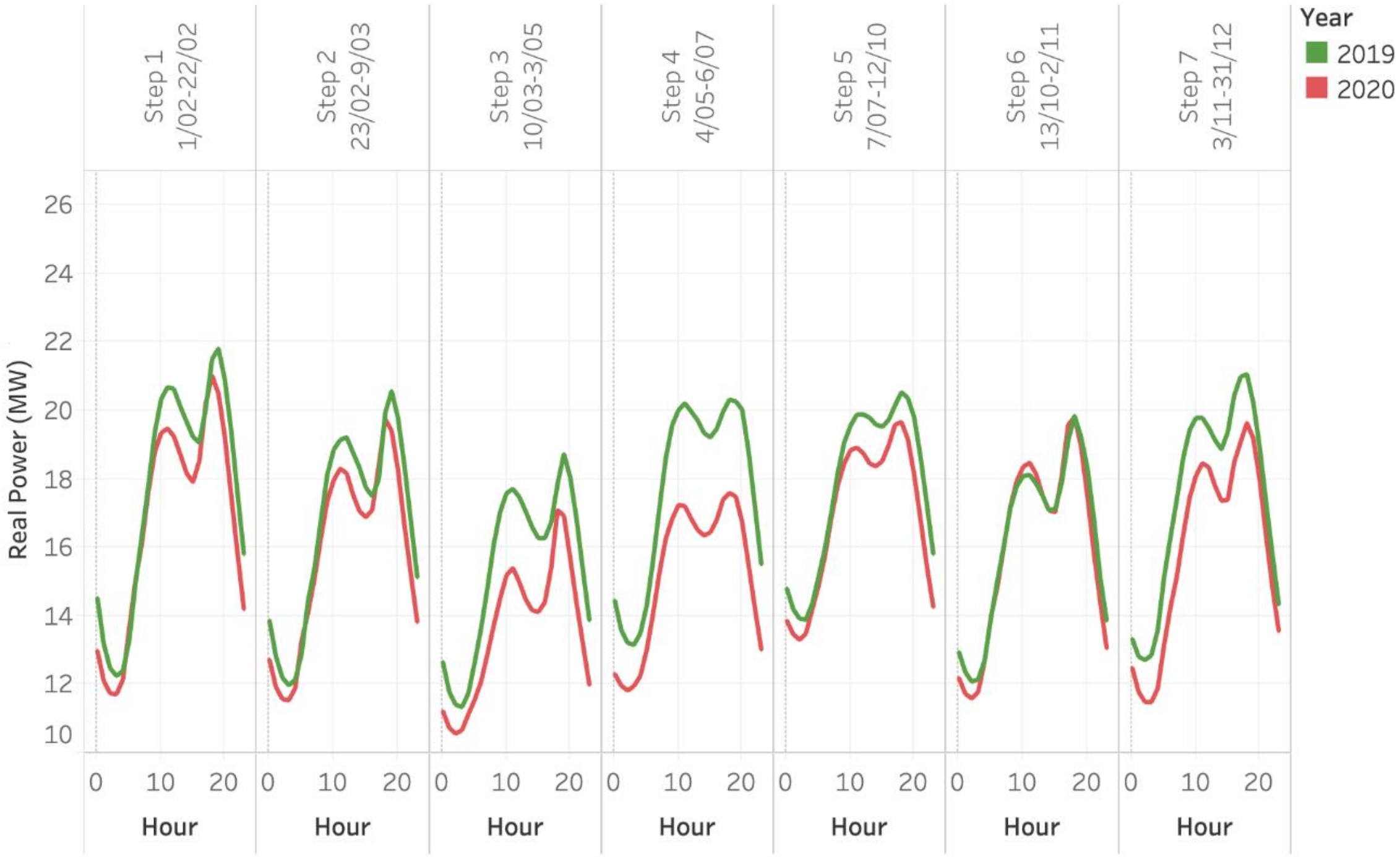

Fig. 6 Weekend real power profiles in the seven steps for 2019 (in green) and 2020 (red line).

It is worth noticing that the trends reported in Fig. 5 and Fig. 6 are computed as the hourly average of the data related to the step, from Monday to Friday for business days and concerning Saturday and Sunday for the weekend.

The changing demand and load profile shape recorded in Milan well pictured the national effect of COVID-19 on the power system operation and the electricity market. As declared in the 2020 Annual Report of the Electricity Market Operator (GME) [28], the exceptional economic situation resulting from the pandemic is also reverberating on the Italian MGP (day-ahead market), causing, among others: i) a drop in prices and volumes, which have fallen to their historical low, compared to liquidity which is, on the contrary, at its greatest; ii) a reduction in zonal differentials, iii) slight changes of market shares by source, to the advantage of renewable generation, iv) an unprecedented level of alignment of quotations at the border, with a consequent reduction to an all-time minimum in the net trade balance with foreign countries. These dynamics reach their maximum intensity in the central part of the year, affected by the first wave of COVID-19 and by highly restrictive measures adopted at the national level for its containment.

Regarding volumes and liquidity, the effects of COVID-19 on the electricity system emerge strongly in the data relating to the energy demand measured by Terna, the Italian TSO, which fell to 302.8 TWh (-5.3%). To better understand the dimensions of this shock, it is enough to observe that one must go back to 2000 to find a lower consumption. Moreover, annual reductions of this magnitude were recorded exclusively in 2009, the last intense economic crisis, and the second post-war period. Within this scenario, the overall trade on the MGP also reached an all-time low of 280.2 TWh (-5.5%). The decreases mainly affect the period March-July, in which about 85% of the overall annual decline is concentrated. Opposite dynamics for liquidity, which instead reaches the highest value ever (74.9%, +2.8 percentual points). As a consequence, the PUN (single national price) reaches the lowest value ever recorded since the start of the electricity exchange, equal to 38.92 €/MWh (-13.41 €/MWh, -25.6%), following a dynamic that i) is common to all the primary European electricity quotations; ii) reflects the significant reductions in market volumes and the cost of the gas raw material (10.55 €/MWh, -35.2%); is it related to the high availability of renewable offer; iii) compresses the "clean spark spread" on values lower than 2019 (9.7 €/MWh, -22.1%) - but similar to 2018 - effectively eliminating it between April and May; iv) it happens throughout the year, assuming considerable intensity between January and August. Concerning this last point, it is interesting to highlight that, starting from March, the share of variation of the PUN does not explain by generation costs increases, differently from the previous months of January and February when the price trend appeared almost entirely attributable to generation costs. A historical minimum level for the PUN has also been recorded in the individual groups of hours, for a peak/off-peak work ratio that stands at 1.2 on an annual basis (+0.03), and which records reversals in April and May, resulting slightly lower than the unit as only sporadically happened in the past. The peculiar dynamics observed during 2020, as well as their increased interdependence on a European basis, show their impacts on the variability and the minimum and maximum hourly values of the PUN, favoring: i) a return of volatility close to the highest level of the last decade (12%), especially between March and June (18%-22%); ii) hourly minima at 0 €/MWh in 5 hours (in April), occurred only twice in the past; iii) an hourly peak of 162.57 €/MWh recorded at 8 pm of September 15$^{th}$ 2020 as a reflection of mainly extranational dynamics, located in France (increase in demand, low level of nuclear production) and Germany (low level of wind production), at the same time characterized by quotations at 190 €/MWh.

The dynamics observed on the MI (Intraday Market) confirm their close relationship with the day-ahead market, showing, in particular, a consolidation of the support function carried out by this market for the definition of efficient programming of the plants. In this sense, the propensity to trade in sessions close to real-time strengthens, with a simultaneous progressive increase in the price range with the MGP. In the emergency context that characterized 2020, these elements emerged with particular importance in the central part of the year, when the uncertainty induced by the first phase of the health emergency accentuated the need to postpone as much as possible the adjustment of the programs.

Concerning the Ancillary Services Market (MSD), on April 7$^{th}$, 2020, the Italian Regulatory Authority for Energy, Networks, and Environment (ARERA) published the technical report titled "temporary economic enhancement of imbalances in the presence of the epidemiological emergency from COVID-19" [29]. The report regards the resolution 121/2020/R/eel [30], which temporarily modified the current regulation to determine imbalance fees considering the effects of the epidemiological emergency. The resolution tried to face the effects of the suspension of activities throughout the national territory which causes a significant reduction in electricity consumption and, therefore, a reduction in the prices of the sales offers accepted on the MGP, and an increase in the difficulty of programming the dispatching points in withdrawal, which causes a greater overall imbalance cost. In addition, the significant reduction in electricity consumption in a context characterized by a non-negligible production of electricity from non-programmable renewable sources may lead to more significant difficulties in the safe operation of the national electricity system. Moreover, in this situation, it appears that, in the MSD, for real-time balancing, purchase offers or offers for sale with prices significantly different from the prices that are formed on the MGP in the same period, offers that can be associated with movements other than those necessary to compensate for the actual imbalances. Therefore, the resolution introduces elements that allow limiting the variability of the imbalance price, also for the prices of the sales offers accepted on the MGP while maintaining adherence to the service costs as far as possible. In particular, to calculate the imbalance prices to be applied to dispatching units not compulsorily enabled, the prices of the offers to buy or sell accepted on the MSD shall be modified so that they fall within a range between a minimum value and a maximum value, defined on a conventional basis. The intervention has been applied from March 10$^{th}$, 2020, i.e., from the day the first measures to contain the COVID-19 epidemic that has significantly affected electricity consumption1were effective throughout the national territory.

Finally, regarding the power generation, the reduction in purchases of 2020 is entirely reflected in sales of thermoelectric plants (140.5 TWh, -8.8%), also due to the growth of renewables, whose volumes are lower only than the maximum of 2014 (95.9 TWh, +0.4%). For traditional sources, the most significant impacts in percentage terms concern coal (7.1 TWh, -47.9%), also in correspondence with emission costs practically aligned with the maximum of 2019, and other traditional thermals (13.3 TWh, -14.2%), while gas-fired plants maintain their volumes lower in the last nine years only than those of 2019 (120.1 TWh, -5.3%), partially offsetting the decline in the external balance. Among renewables, the growth of hydroelectric generation (+1.6%), concentrated in the North in the first five months of the year, and that of solar (+5.7%) stand out, which counterbalance the decrease in wind volumes (-8.1%), lower however only to the maximum of last year. In terms of mix, the share of the gas market remained at around 43%, while that of renewable sources rose to 34% (+4 p.p.), half of which was the prerogative of hydroelectric plants (17.2%, +1.2 p.p.), pushed by the results recorded between April and June (maximum peak at 48% in May). The contribution of imports to cover national needs was substantially stable at around 15% but fell between April and June to shallow values between 7% and 10%.

## *B. Reactive Power*

The changes in people's lifestyles have also reverberated on the distribution network's reactive power. Fig. 7 shows the average HV/MV transformer's reactive power trend in 2019 and 2020. As for real power, seven vertical lines are shown to mark starting and ending dates of the steps defined in Section II. Moreover, the monthly average reactive power is drawn to highlight the changes recorded in 2020 compared to 2019. The reactive power at the point of common coupling between distribution and transmission networks depends on several contributions: the reactive power either withdrawn or injected by the customers, the reactive power consumed by the cable's inductance, and the reactive power produced by the cable's capacitance. Since the distribution network of Milano is mainly made of underground cables, which substantially contributes to producing reactive power at the distribution level, in the following, we concentrate on the reactive reverse power flow.

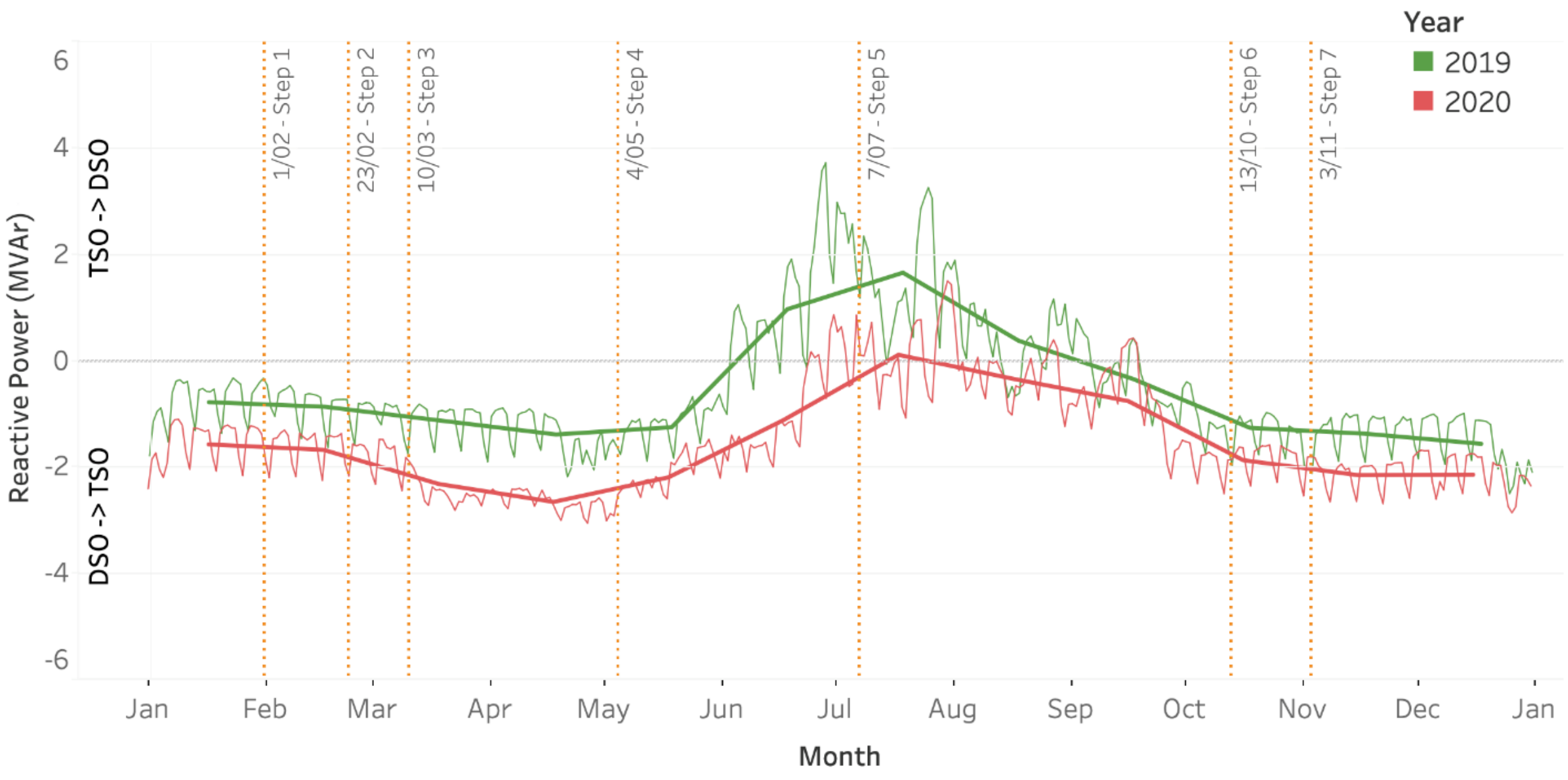

Fig. 7 Average HV/MV transformer's reactive power in 2019 and 2020.

Similar to what was found for active power, on the one hand, the 2019 curve has a layout of five business days and two lower weekend days. Moreover, the same seasonal trend clearly shows up: in wintertime, the power demand is slightly higher than in spring and autumn; due to the usage of air conditioners, the reactive demand starting to increase from June, reaching the peak at the end of June; a valley in the middle of August identified the usual summer vacations. It is worth noticing that the reactive power value is always negative, i.e., there is a reverse reactive power flow from the distribution to the transmission network, apart from the summer period, from June to September. Out of the summer, the capacitive contribution of the power cables overcompensates the reactive request of customers and cable's inductances. On the other hand, the 2020 load profile has been affected by the consequence of COVID-19. The comparison between the two years is better outlined in Fig. 8. Fig. 8 reports the reactive power values and their variation. The middle graph considers the reactive power variation while the lower one the absolute value. As a general consideration, it is evident that, in 2020, the flow of reactive power increases; in particular, the reverse reactive power flow is more predominant. Only in a few days of July, the average reactive power is positive, i.e., it is flowing from the transmission to the distribution network. Moreover, it can be noticed that:

- In the first months of 2020, it is evident a significant increase of reverse reactive power flow, which increase around 100% in Step 1 (1/02 - 22/02), Step 2 (23/02 - 9/03), and Step 3 (10/03 - 3/05). The reactive power flowing on the distribution network increases, reaching its yearly maximum in Step 3;
- Like the active power, the highest variation is recorded in Step 4 (4/05 - 6/07), where the absolute variation between the two years is more than 150%. It is worth noticing that, apart from the considerable variation, from June to July, no reactive reverse power flow was measured in 2019 so that the net value in the step results positive, while in 2020, the reactive power flows from the distribution to the transmission network. The behavior is highlighted by the considerable variation, more than 300%, of the corresponding bar depicted in the middle graph. Starting from May, the slope of the curves increases, but the reactive reverse power flow remains, except for a few days of July. The increasing trend in both years is due to the highest reactive demand linked to the use of air conditioning systems which compensate, in 2019 overcompensates, the reactive power produced by the cable's capacitance;

- In Step 5 (7/07 - 12/10), the reactive power at the point of common coupling is reduced, on average, by 50%. The reduction is concentrated in July, while September and August have more similar values to 2019. As for Step 4 (4/05 - 6/07), it is worth noticing that, apart from the variation, from July to September, no reactive reverse power flow was measured in 2019 so that the net value in the step results positive, no reactive reverse power flow was measured in 2019, while in 2020 is recorded.
- In the remaining months, Step 6 (13/10 - 2/11) and Step 7 (3/11 - 31/12), the variation between the two years becomes lower, with a reverse power flow that increased within 25% and 50% of the 2019 ones.

By comparing Fig. 4 with Fig. 8, it can be seen that there is a greater increase of the reverse reactive power flow with respect to active power variations. On the one hand, the capacitive contribution of cables, which depends on the voltage square, could be considered almost constant. However, on the other hand, reducing the power demand lessens the reactive power withdrawn by the customers and, consequently, the reactive power consumed by the cable's inductance. Since the latter contribution depends on the current square, its reduction changes more significantly than the reduction in power demand. Moreover, the overamplification of the reverse reactive power flow could also highlight a change in the load types, from industrial/commercial to residential, which typically have a less inductive power factor. To better understand this active-reactive behavior, an example of Boucherot's theorem applied on a 3-phase medium voltage feeder is reported in Appendix B.

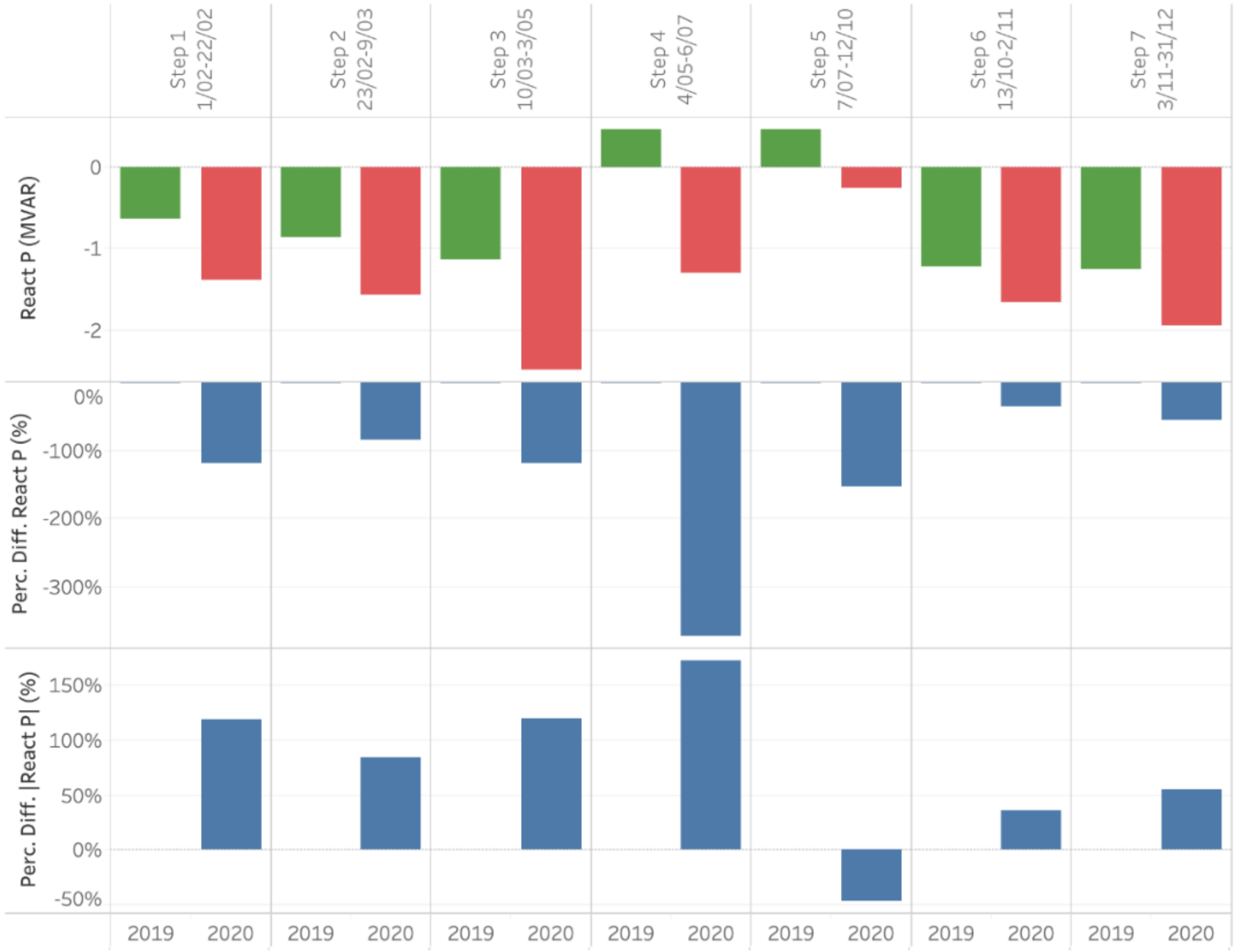

Fig. 8 Average reactive power measured in each of the seven steps. 2019: green; 2020: red. In blue is the percentage variation.

Fig. 9 summarizes the relationship between active and reactive power using the power factor. For the sake of clarity, four colors have been used to distinguish power flows from the transmission to the distribution network (Q>0) from power flows from the distribution to the transmission network, i.e., reverse power flow (Q<0). Two are the primary outcomes: the reverse power flow suddenly falls starting from March 10$^{th}$, 2020, while its trend was more stable in 2019; during 2020 summer months, the power factor of the Milano distribution network was almost 1, i.e., the distribution grid has been seen by the national grid almost as a pure resistor.

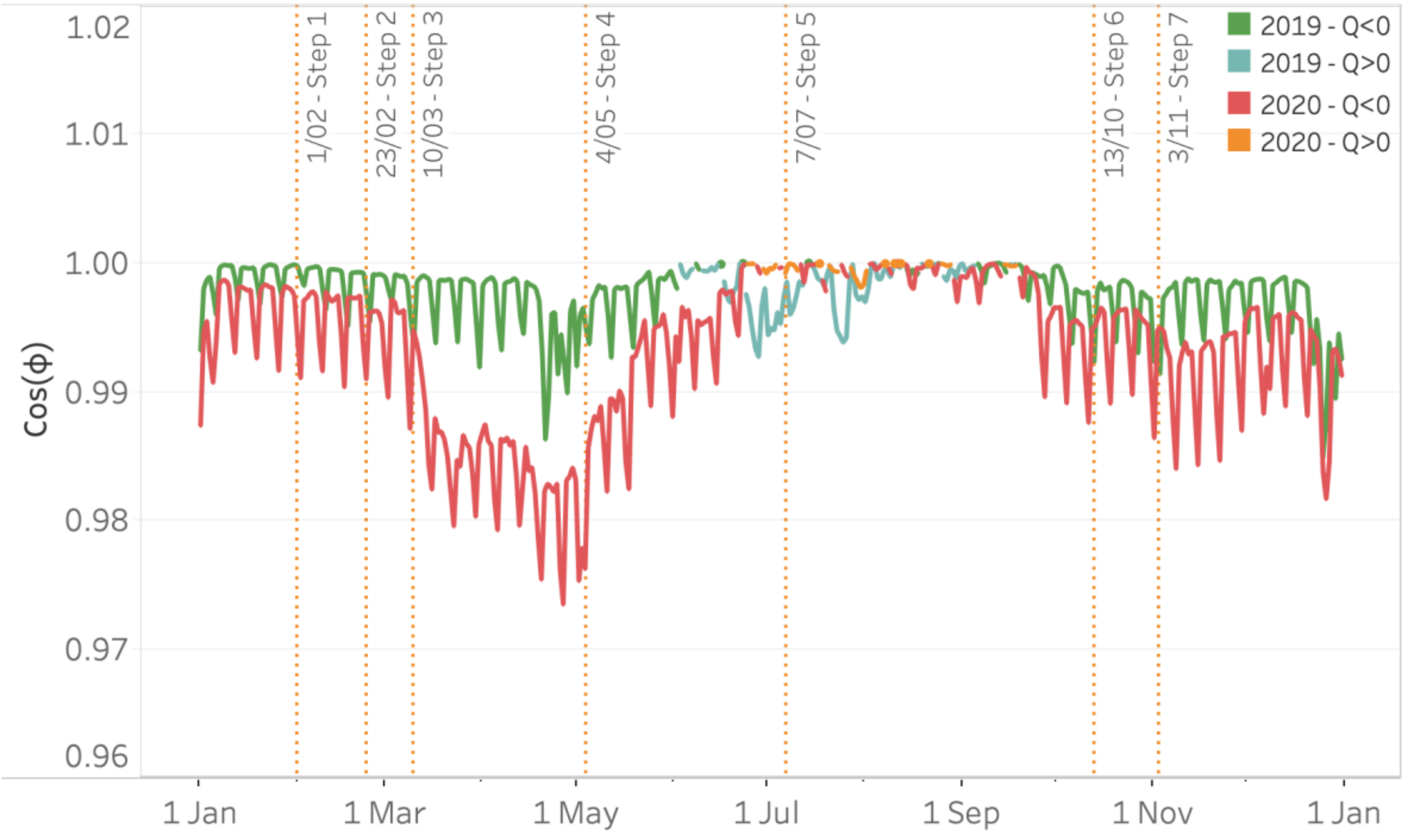

Fig. 9 Average power factor in the seven steps for 2019 (green Q<0 and blue Q>0) and 2020 (red Q<0 and orange Q>0).

As in the case of active power, Fig. 10 and Fig. 11 show the reactive power trend for business days and weekends. However, unlike the active power, in both cases, the shape of the two curves does not show significant changes. Concerning 2020, while in business days, the reactive flow is from the transmission to the distribution network in Step 5 (7/07 - 12/10), this does not happen on weekend days, when only reactive power reverse power flow was recorded. The behavior could confirm that the reduction of the industrial/commercial activities in favor of the residential one amplifies the capacitive behavior of the distribution network, a sign of a more inductive feature of industrial and commercial loads.

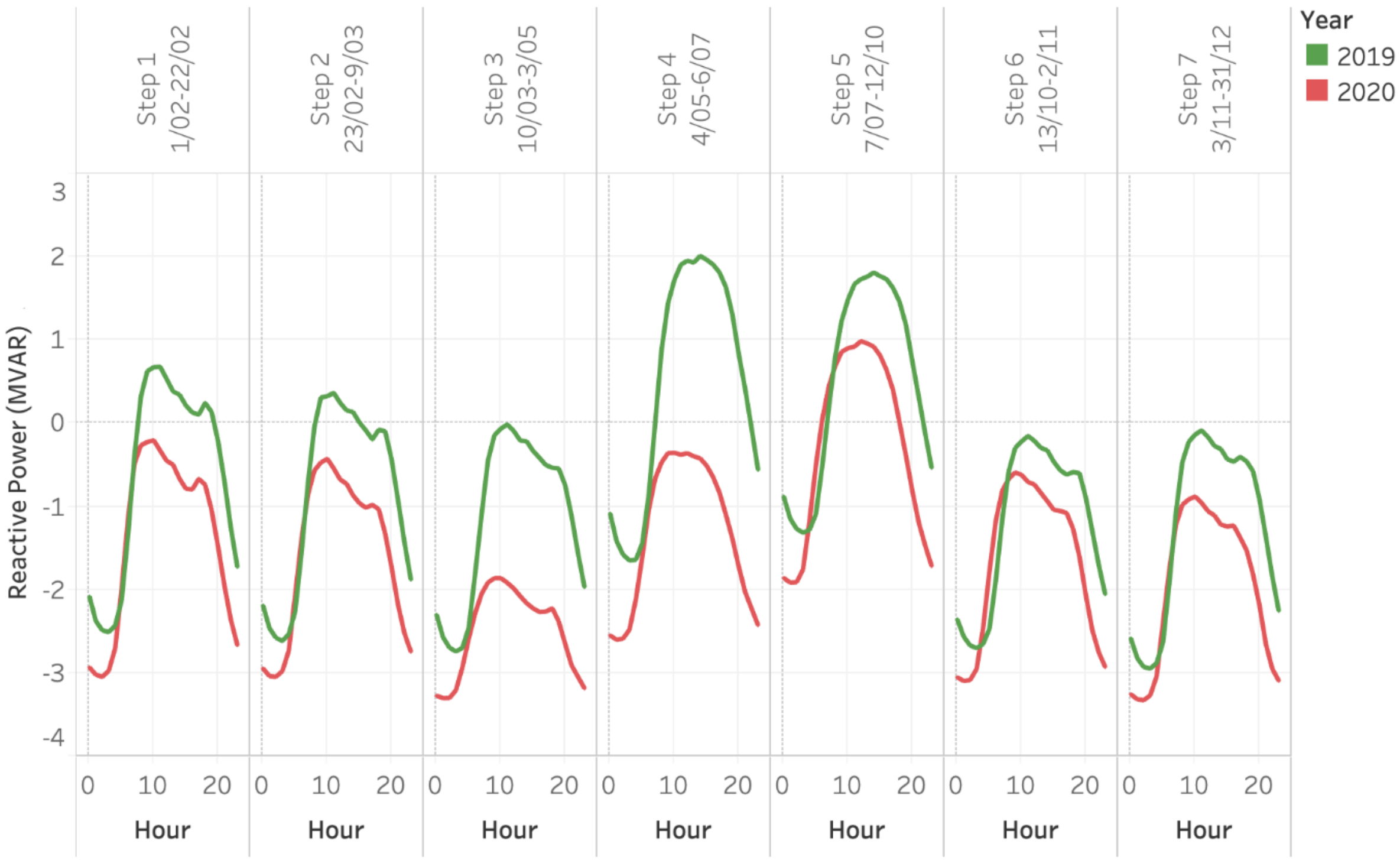

Fig. 10 Business days reactive power profiles in the seven steps for 2019 (in green) and 2020 (red line).

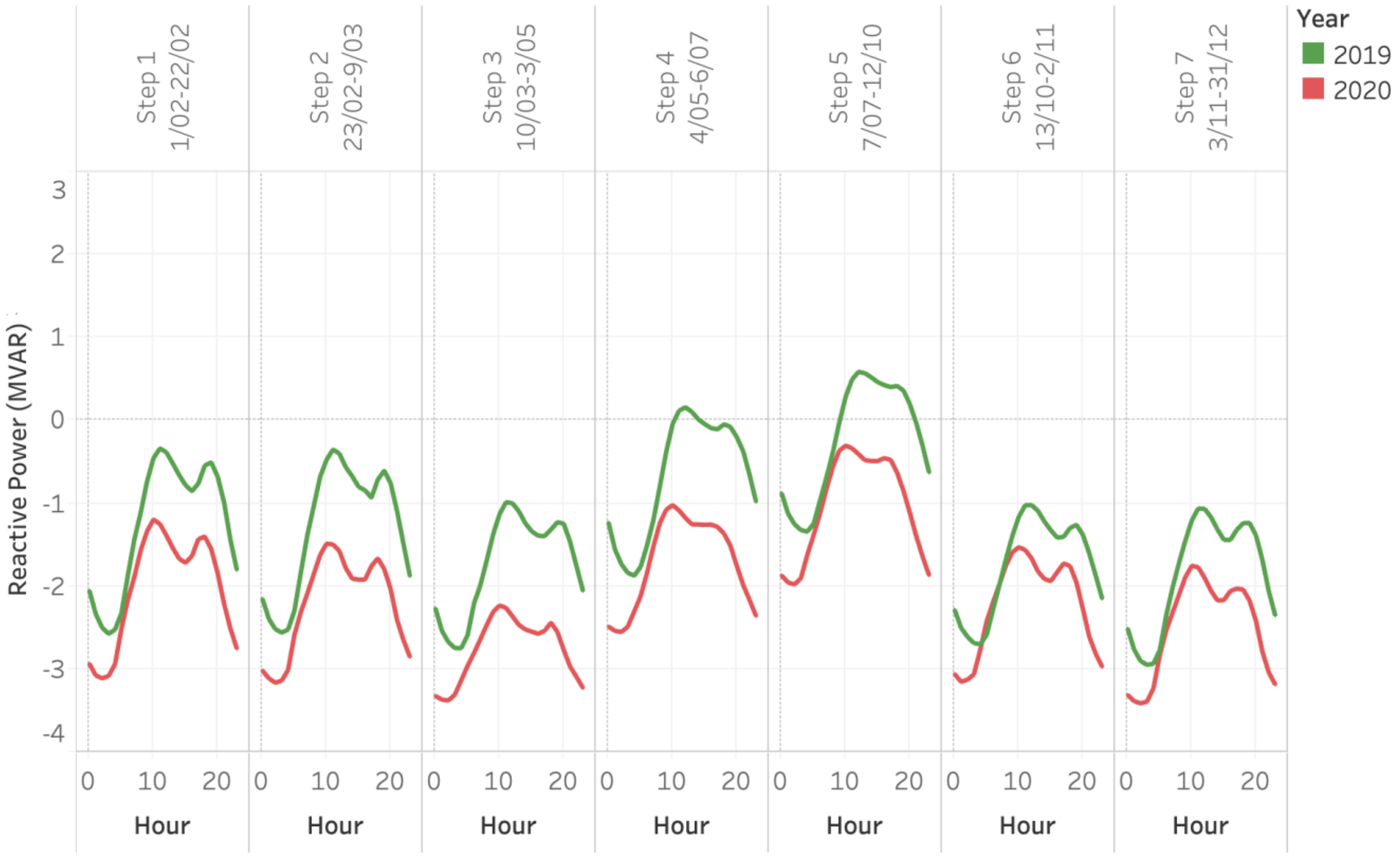

Fig. 11 Weekend reactive power profiles in the seven steps for 2019 (in green) and 2020 (red line).

As for real power, it is worth noticing that the trends reported in Fig. 10 and Fig. 11 are computed as the hourly average of the data related to the step, from Monday to Friday for business days and concerning Saturday and Sunday for the weekend.

Finally, Fig. 12 and Fig. 13 report the relationship between active and reactive power using power factor as a proxy measure for this relation. In 2019, the power factor was exceptionally flat with a unitary value during business days. Its value falls (Q<0) during night hours and in summer months due to the increased load (Q>0). Similarly is for 2020, except for Step 3 (10/03 - 3/05). The evident drops during night hours cause a substantial increase in the average reverse power flow, already shown in Fig. 8.

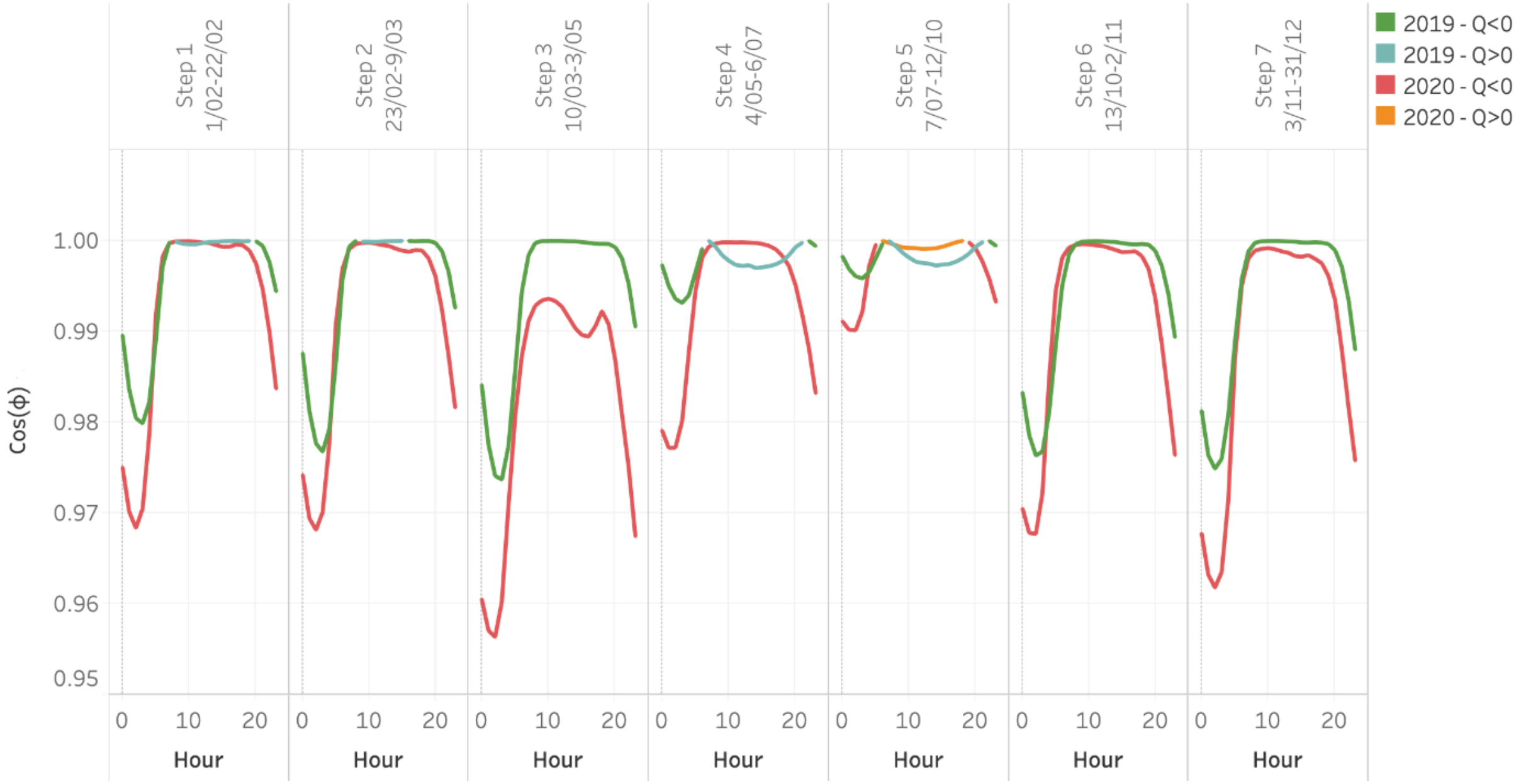

Fig. 12 Business days factor in the seven steps for 2019 (green Q<0 and blue Q>0) and 2020 (red Q<0 and orange Q>0).

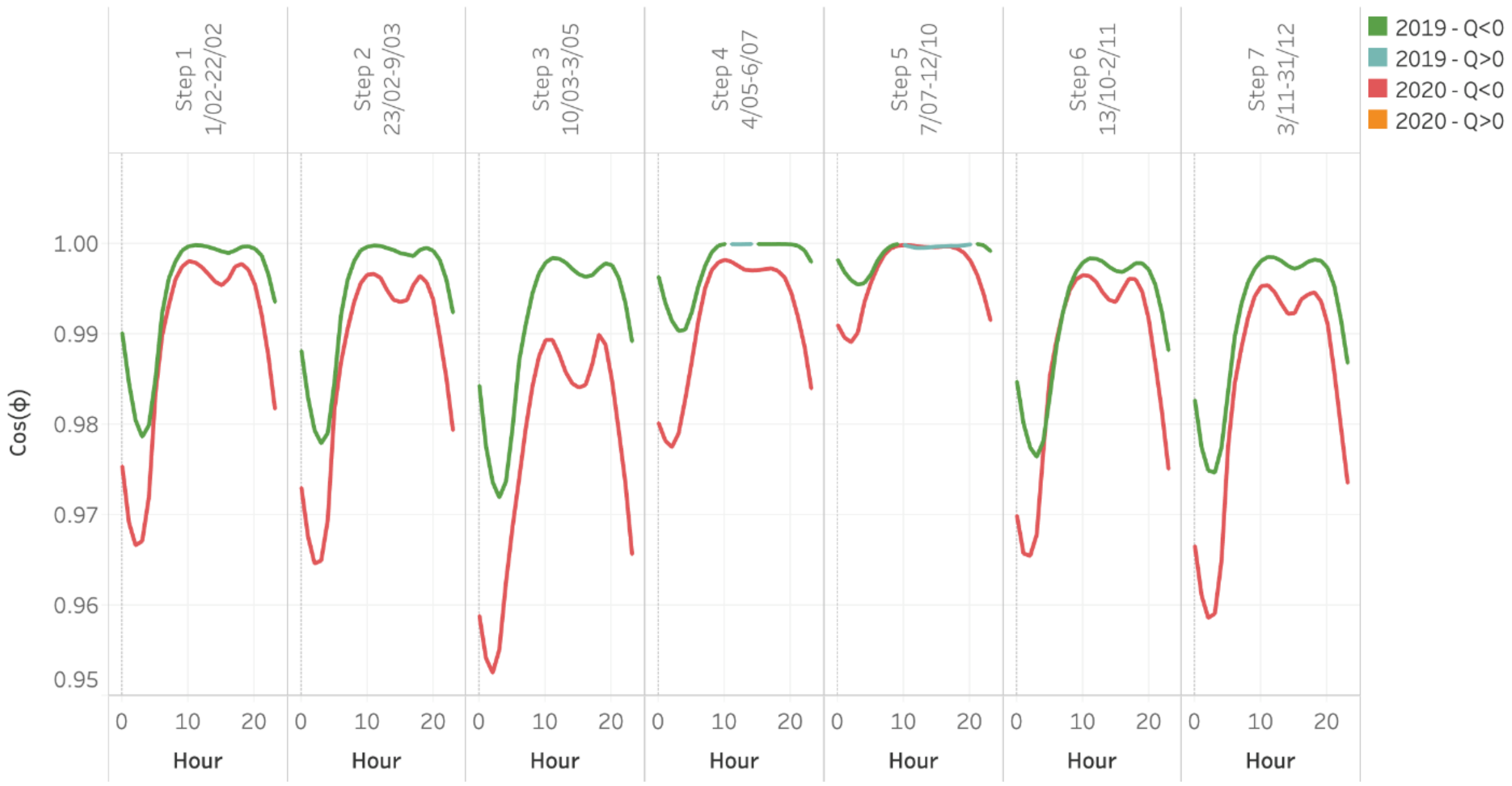

Fig. 13 Weekend factor in the seven steps for 2019 (green Q<0 and blue Q>0) and 2020 (red Q<0 and orange Q>0).

The changing in the reactive power flow at the point of common coupling between high and medium voltage networks has been under the attention of ARERA since 2016. Therefore, the effect of COVID-19 on the reactive reverse power flow well fits this framework. With the consultation paper 420/2016/R/eel "Tariff regulation of reactive energy for high and extra-high voltage networks and distribution networks" [31], ARERA illustrates its guidelines aimed at updating the regulatory framework on reactive energy, concerning the withdrawals and inputs of reactive energy into high and very high voltage networks by end customers and distribution system operators (DSOs), through the introduction of fees as close as possible to the costs. The Regulating Authority considers appropriate changes to make final customers—which include DSOs—responsible for reactive power control. Considering the current regulatory framework, the document calculates the costs incurred by Terna on the MSD to ensure proper network operation according to the flows of reactive energy and voltage constraints in the various nodes. Thus, the document sets out the Authority's general guidelines about the optimal power factor levels. On the one hand, it proposes to modify the minimum power factor level of the reactive energy withdrawals, considering, during daylight hours, a limit power factor of 0.95. On the other hand, it proposes to introduce a minimum power factor of the reactive energy input equal to 1, to limit its input into the network for all daylight and night hours.

Moreover, the document illustrates specific guidelines for distribution companies, providing that inputs and withdrawals of reactive energy by DSO could be controlled through initiatives in terms of: installation of reactive energy control systems in its networks; input to end-customer facilities connected to medium-voltage distribution networks; input to distributed generators connected to medium voltage distribution networks, or internal to end customers' plants, with a nominal power exceeding 100 kW. Furthermore, concerning the inputs of reactive energy by end customers and medium and low voltage distribution companies, the Authority intends to evaluate the possibility of introducing price signals. In this way, the reactive power inputs fees could be defined according to the costs associated with the management of reactive energy in MSD.

The change in reactive power flows can also affect the grid protection concepts, especially distance and overcurrent relay. For instance, paper [32] investigates the effect on the grid protections of the energy transition, where large thermal power plants are gradually being taken off the transmission grid, and a large number of renewable sources are being added to the distribution grids. The paper reports that both the increase and the change of the reactive power flow can lead to false tripping of the protective devices.

## V. COVID-19 LOCKDOWN IMPACT ON FAULTS

Distribution grids reliability is an essential issue for DSOs. In order to increase the quality of service, a Performance-Based Regulation (PBR) was introduced by ARERA in the resolution 646/2015/R/eel (*Integrated text of the output-based regulation of electricity distribution and metering services for the period 2016-2023*) [33]. The PBR relies on the System Average Interruption Frequency Index (SAIFI) and the System Average Duration Frequency Index (SAIDI). SAIFI is calculated considering short (from 1 second to 3 minutes) and long (more than 3 minutes) interruptions under the responsibility of DSOs, while SAIDI only considers long interruptions. The expected goal is to increase the quality of service, stimulating DSOs to be more efficient in their investments. In a PBR regime, rewards and penalties are linked to specific targets. DSOs providing high-quality service will receive awards, while penalties are given to DSOs that provide low-reliability services. SAIFI and SAIDI do not take into account force majeure events. On the one hand, those events are becoming more and more frequent with increased impact in terms of the number of interrupted users and duration. On the other hand, if these events were considered, reliability indices would not reflect the actual reliability of the distribution network. However, it is not easy to identify the minimum conditions to link the interruptions to a natural hazard rather than DSO responsibility. Some criteria for identifying Major Event Days (MEDs) are given in [34] and [35]. The increasing number of MEDs has led the Authority to include the concept of network resilience into the regulatory framework and allocate resources to incentive DSOs in increasing networks resilience.

As shown in the Unareti development and resiliency plan [22], doing an in-depth analysis of the type of fault element in the last five years, the cable joint is the leading cause of failure on the medium voltage distribution network. However, the evidence from the field shows that the criticality is not intrinsic in the cable joint but instead in the coupling between the joint and the cable insulation layer, especially for the joints that combine different types of cables. The presented analysis investigates whether COVID-19 pandemics affected the outages occurrence in 2020. In particular, the investigation focuses on the number of interruptions, which is historically the most affecting the quality of service for the Milan distribution network. Fig. 14 shows the number of faults on the MV Unareti grid in 2019 and 2020 and the monthly maximum fault range in 2012-2020. Apart from the summer months, looking at Fig. 15, which focuses on April and May, it is worth noting that the highest number of failures was recorded in 2020. However, these numbers are still of the same order of magnitude as the previous years and do not justify a particular effect linked to the COVID-19 pandemic. As reported in [22], Unareti already includes in his network development plan several investments to increase the reliability of the Milan distribution network. The main planned projects are the construction of new primary substations; the increase in the number of remote controls and network automation; the renewal and development of the MV ad LV networks; the installation of Petersen coils in primary substations; the renewal of protection systems; the preventive diagnostics of MV cables; the renewal of secondary substations. All the mentioned actions will positively affect the future quality of service.

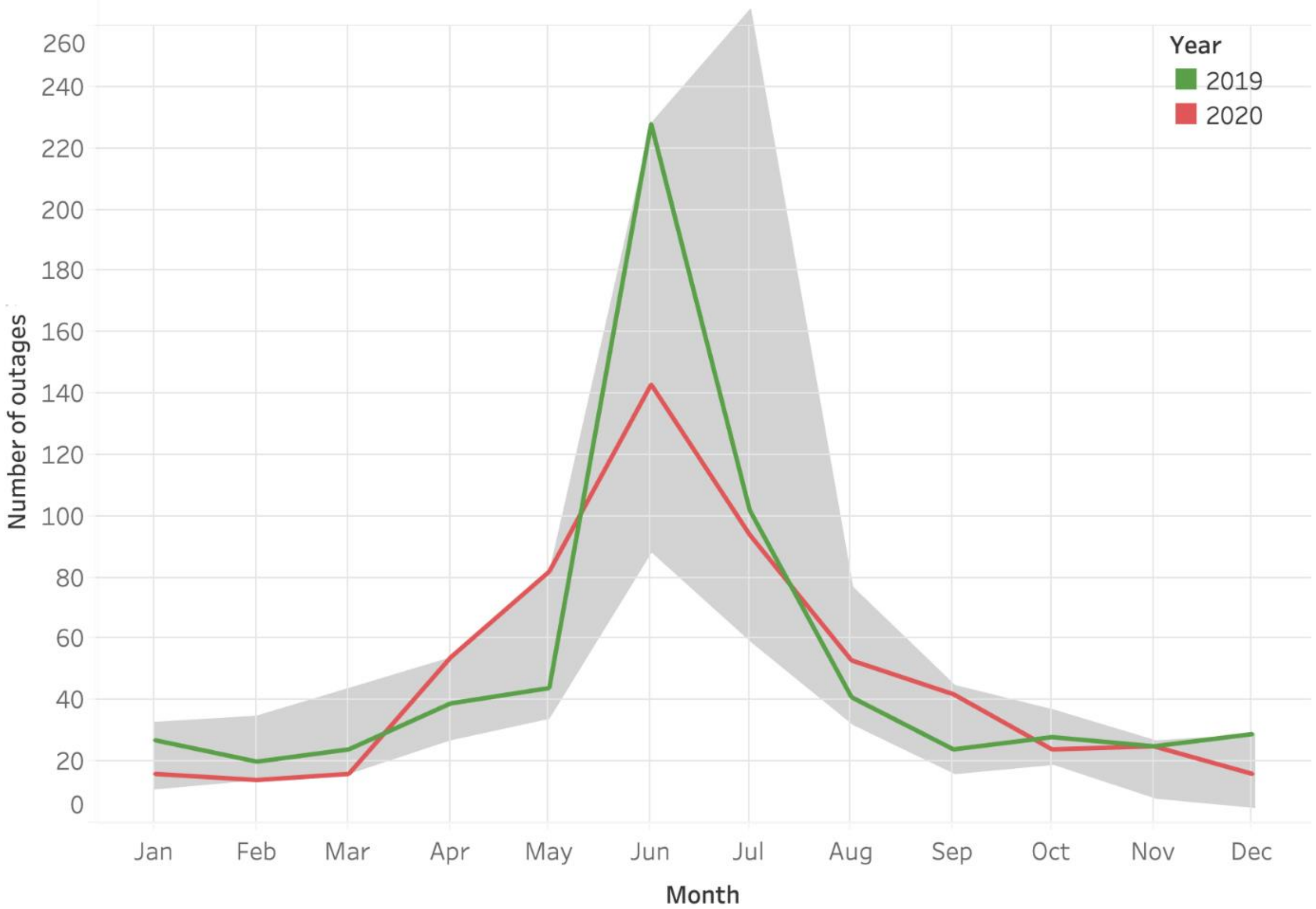

Fig. 14. Number of outages in each month of 2019 (in green) and 2020 (in red) on the Milan Unareti distribution network. The grey band shows the minimum-maximum range of outages within the period 2012-2020.

A different approach has to be used for comparing the summer months. The data reported in Fig. 14 and Fig. 15 show a substantial failure increase in June and July, with significantly more faults in 2019 with respect to 2020. As reported in [22], the events of most significant impact on the Milan Unareti electricity distribution network reliability and resilience correspond to situations of intense and persistent heat (heat waves) and exceptional rainfall (water bombs). During heat waves, which typically occur in the summer months, the network components work under stress. On the one hand, the increase in ambient and soil temperature is directly reflected in the electrical components. On the other hand, additional components heating is related to the increase in the energy demand, the consequence of the massive use of air conditioners. The phenomenon of the components overheating is amplified by the fact that the temperatures remain outside the values of the seasonal averages even during the night hours for several consecutive days. Comparing Fig. 2 and Fig. 3, during the 2019 heatwave, there was an increase in the electrical load correlated with the increase in the ambient temperature. As shown on the right-hand side of Fig. 15, in 2019, the phenomenon reflected in the number of faults: the exceptional number of failures in June 2019 is, therefore, linked to the intense heatwave that affected Milan between 26$^{th}$ and 29$^{th}$ June 2019.

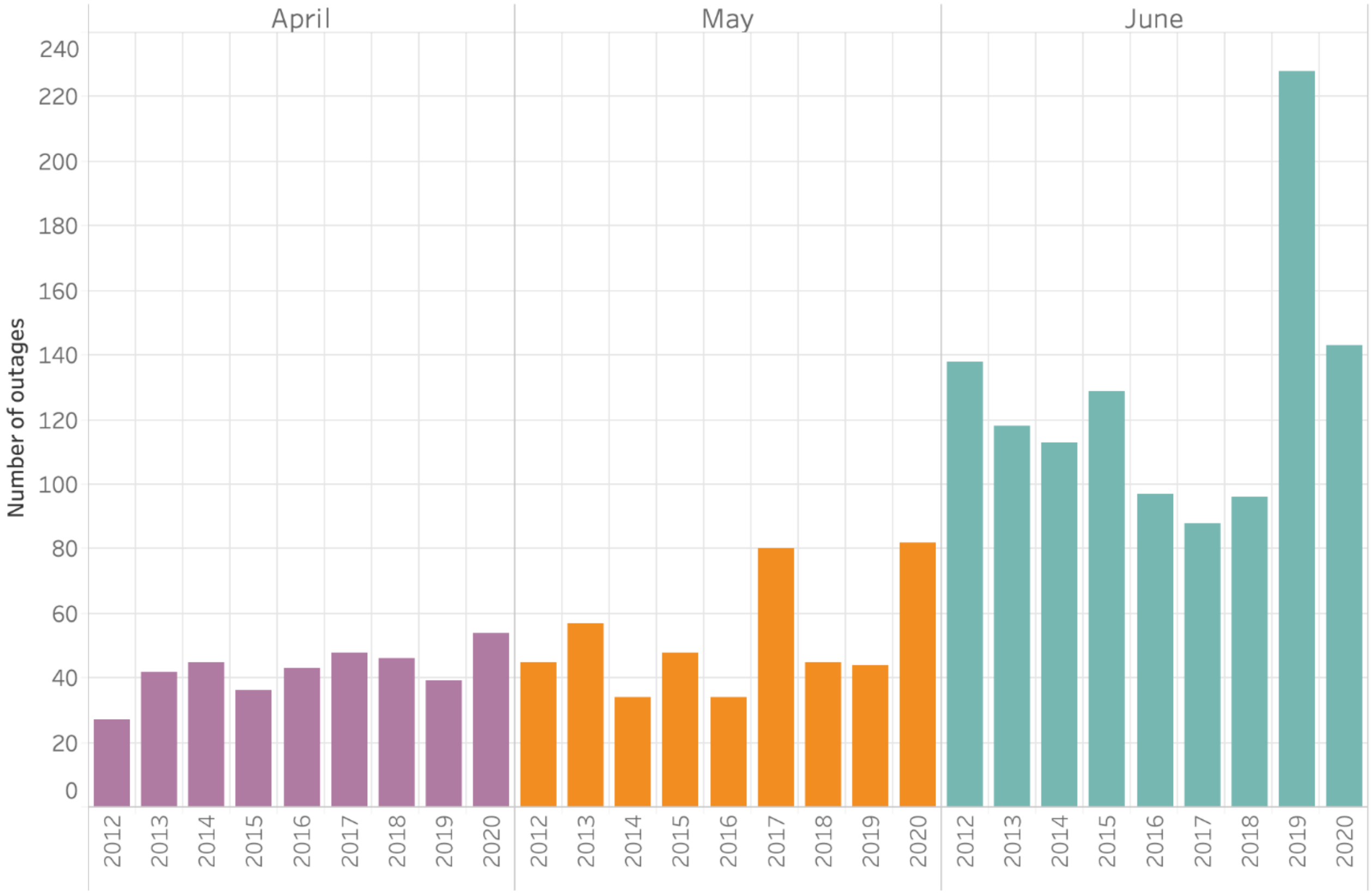

Fig. 15. Number of outages in April, May, and June on the Milan Unareti distribution network. Data are available for the years 2012-2020.

To better investigate the relationship between temperature increase and the number of failures, in Fig. 16, the temperature is divided into 21 bins, from -2°C to 39 °C. The upper graph represents the temperature frequency per bin, the middle graph the number of outages begun within the temperature bin, and the lower graph shows the percentages of outages begun within the temperature bin, out of the total number of temperature samples in the bin. It is evident that the outage percentages increase with increasing temperatures and that above the threshold of 34°C, the percentage of beginning outages overcomes 2%. However, this threshold is never reached in 2020, while in 2019, it clearly identifies the June heatwave. All considered, the analyzed data do not justify a particular effect linked to the COVID-19 pandemics also during the summer months.

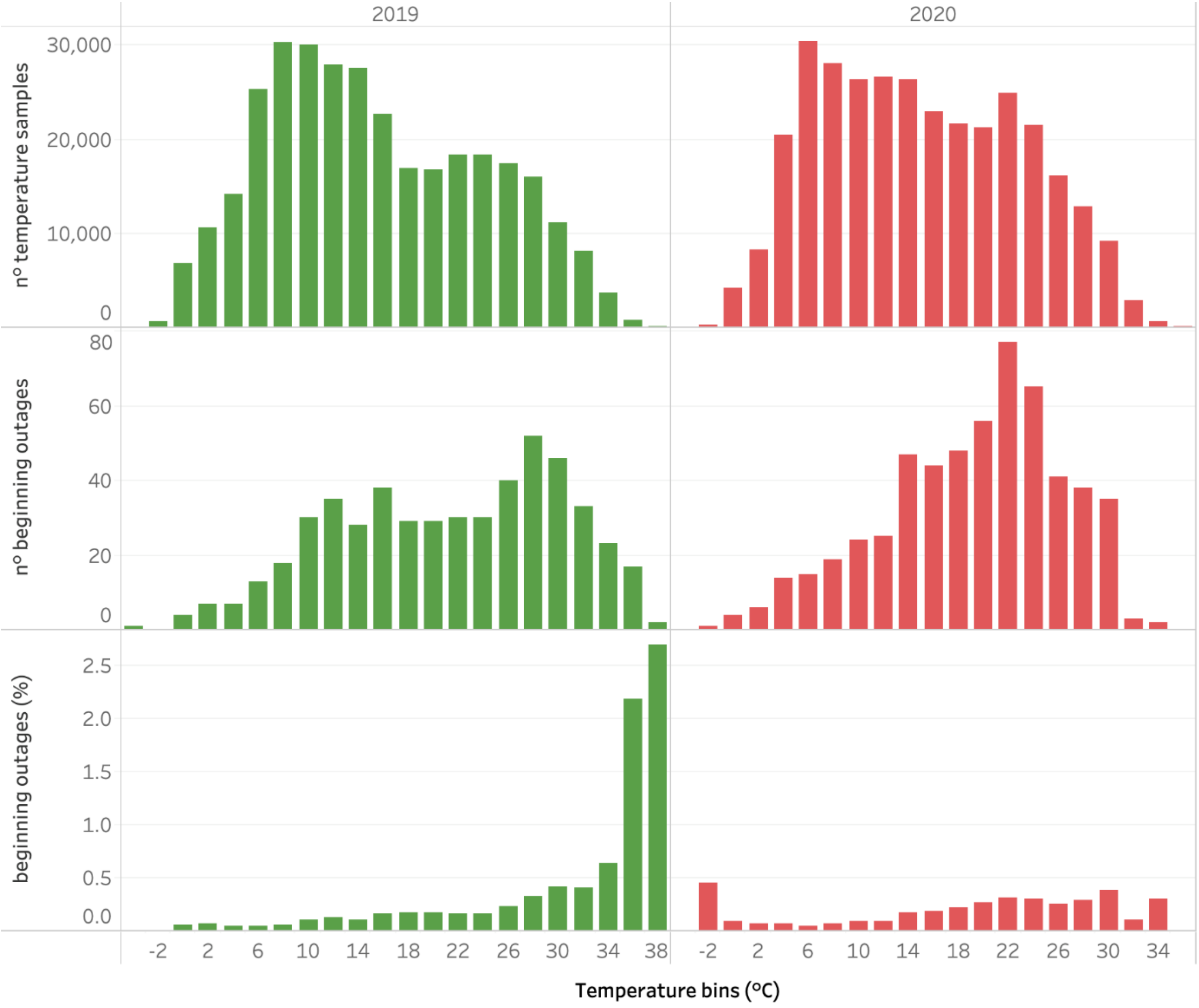

Fig. 16. Upper graph: number of temperature samples per temperature bin. Middle graph: number of outages begun within the temperature bin. Lower graph: percentages of outages begun within the temperature bin over the total number of temperature samples in the bin.

## VI. Conclusion

The paper presents the lessons learned from analyzing the Milan distribution networks data during the COVID-19 pandemic. The analysis considers active and reactive powers and faults information comparing 2019 and 2020. On the one hand, regarding active power, the Milan distribution networks recorded a reduction of the power demand within the range of 5%-20% and a change in the load profile shape. March and April 2020 load profiles show a pattern with the highest peak in the evening, the shape of a typical residential user, confirming the closure of most industrial/commercial activities and the more prominent presence of people at home. Thus, a shift to a more diffuse work from home brings power demand down and changes the daily load to a more residential-related trend. Consequently, the day-ahead market would be affected by a drop in prices and volumes, a reduction in zonal differentials, slight changes of market shares by source, to the advantage of renewable generation, an unprecedented level of alignment of quotations at the border, with a consequent reduction to an all-time minimum in the net trade balance with foreign countries. It can be assumed that an increase in the energy demand for electrification of end-users may have the opposite effects.

On the other hand, the Milan distribution network recorded a higher flow of reactive power. In the first semester of 2020, the variation is within the range of 75%, 150%. Moreover, a more predominant reactive reverse power flow was recorded at the point common of coupling between the transmission and distribution network. Both these consistent changes are a consequence of the capacitive nature of the Milan MV distribution cables mixed with a reduction in the user's reactive power demand. The capacitive behavior is also highlighted by the leading power factor predominantly recorded in 2020. In urban underground distribution networks, active power reduction greatly impacts reactive power flows: the more the active power reduction, the more the reactive reverse power flow from the DSO to the TSO. The changing of reactive power flows also reflects on the ancillary service market, leading to additional costs to ensure proper network operation. Again, it can be assumed that an increase in the energy demand for electrification of end-users may have the opposite effects, reducing or even inverting the reverse reactive power flows.

Finally, the paper investigates whether COVID-19 pandemics affected the outages in 2020. Based on the analysis's outcomes, the recorded faults in 2020 are of the same order of magnitude as the previous years, so no link to the effects of the COVID-19 pandemic shows up. However, it is worth noting that an increase in energy demand for end-user electrification can have serious consequences on the reliability and resilience of distribution grids, which must be managed primarily with heavy investments in grid infrastructure.

## APPENDIX

### *A. Input data*

TABLE II. PERCENTAGE OF MISSING ELECTRICAL DATA AND NUMBER OF RECORDED OUTAGES

| **HV/MV Transformer** | **Missing Real Power data** | | **Missing Reactive power data** | | **Number of outages** | |
|---|---|---|---|---|---|---|
| | ***2019*** | ***2020*** | ***2019*** | ***2020*** | ***2019*** | ***2020*** |
| PS1_TR1 | 1 | 2 | 1 | 2 | 20 | 12 |
| PS1_TR2 | 1 | 38 | 1 | 38 | 1 | 11 |
| PS1_TR3 | 12 | 2 | 12 | 2 | 22 | 10 |
| PS2_TR1 | 63 | 2 | 0 | 2 | 10 | 20 |
| PS2_TR2 | 4 | 3 | 4 | 3 | 23 | 39 |
| PS2_TR3 | 1 | 12 | 1 | 12 | 63 | 47 |
| PS2_TR4 | 81 | 56 | 75 | 56 | 1 | 0 |
| PS2_TR5 | 18 | 44 | 19 | 44 | 0 | 0 |
| PS3_TR1 | 0 | 0 | 0 | 0 | 2 | 10 |
| PS3_TR2 | 0 | 0 | 0 | 0 | 18 | 9 |
| PS4_TR1 | 0 | 0 | 0 | 0 | 0 | 0 |
| PS4_TR2 | 0 | 2 | 0 | 2 | 46 | 70 |
| PS4_TR3 | 3 | 3 | 3 | 3 | 47 | 52 |
| PS4_TR4 | 0 | 0 | 0 | 0 | 18 | 53 |
| PS5_TR1 | 0 | 0 | 0 | 0 | 55 | 31 |
| PS5_TR2 | 0 | 0 | 0 | 0 | 22 | 13 |
| PS5_TR3 | 0 | 0 | 0 | 0 | 32 | 24 |
| PS5_TR4 | 0 | 1 | 0 | 1 | 27 | 11 |
| PS5_TR5 | 0 | 0 | 0 | 0 | 15 | 20 |
| PS6_TR1 | 0 | 0 | 0 | 0 | 21 | 16 |
| PS6_TR2 | 0 | 0 | 0 | 0 | 53 | 46 |
| PS6_TR3 | 0 | 0 | 0 | 0 | 35 | 38 |
| PS7_TR1 | 0 | 1 | 0 | 1 | 30 | 13 |
| PS7_TR2 | 1 | 2 | 1 | 2 | 20 | 37 |
| PS7_TR3 | 0 | 0 | 0 | 0 | 25 | 33 |
| PS7_TR4 | 3 | 2 | 3 | 2 | 30 | 31 |
| PS8_TR1 | 0 | 0 | 0 | 0 | 4 | 28 |

| | | | | | | | |
|---|---|---|---|---|---|---|---|
| PS8_TR2 | 6 | 0 | 5 | 0 | | 5 | 8 |
| PS9_TR1 | 1 | 0 | 0 | 0 | | 35 | 39 |
| PS9_TR2 | 1 | 5 | 2 | 5 | | 2 | 2 |
| PS9_TR3 | 0 | 1 | 0 | 1 | | 73 | 56 |
| PS9_TR4 | 0 | 0 | 0 | 0 | | 52 | 98 |
| PS10_TR1 | 0 | 1 | 0 | 1 | | 22 | 13 |
| PS10_TR2 | 0 | 3 | 0 | 3 | | 10 | 19 |
| PS11_TR1 | 3 | 0 | 3 | 0 | | 25 | 36 |
| PS11_TR2 | 0 | 0 | 0 | 0 | | 10 | 8 |
| PS11_TR3 | 7 | 6 | 7 | 6 | | 20 | 21 |

### B. *Boucherot's theorem applied on a 3-phase medium voltage feeder*

To better understand the active-reactive behavior presented in Section IV, let consider studying the simplified 1-phase representation of a 3-phase medium voltage feeder shown in Fig. 17 using the Boucherot's theorem. The models of cables selected include the PI-type equivalent model of lumped parameters. Section D is the MV side of HV/MV transformer, while the feeder's distributed load is considered a lumped load connected at the end of the feeder, corresponding to section A.

Medium voltage feeders are radial lines and therefore are relatively easy to work with by applying Boucherot's theorem. The Boucherot's theorem allows the resolution of AC circuits, imposing the balance of powers between sources and loads. Considering the power flows on the connecting lines between the generator and loads, we can define a section in the circuit, represented by a red line in Fig. 17, where the transit of power coming from the loads must be balanced from the one from the generators. Therefore:

$$\sum_{k=1}^{N_S} P_k = \sum_{n=1}^{N_L} P_n$$

$$\sum_{k=1}^{N_S} Q_k = \sum_{n=1}^{N_L} Q_n$$

The balance is valid in terms of complex power S and in terms of active power P and reactive power Q. Reactive power can have a positive (inductive) sign or a negative sign (capacitive). This fact is significant because the term to the right-hand side $\sum_{n=1}^{N_L} Q_n$ could be equal to zero even if the individual terms are not due to the compensation of positive reactive powers (inductive) with negative reactive powers (capacitive). However, the same thing is not possible for the active powers of the loads that are always expressed by positive numbers. One of the advantages of the Boucherot's theorem is that the electrical quantities are the root-mean-square value of phasor quantities, making the computations easier.

Considering Fig. 17 and the following assumptions:

- The system is 3-phase, symmetrical, and balanced;
- The load at the end of the line is a R–L load working at nominal voltage.

Let us apply Boucherot's theorem considering: $P_L = 2\,MW$; $\cos\varphi_L = 0.95$; $V_L = 23$ kV; r = 0.17 ohm/km; $x_L = 0.15$ ohm/km; c = 0.15 µF/km; g = 0; $L = 5\,km$; $f = 50\,Hz$. $P_L$ is the power demand of lumped load L; $\cos\varphi_L$ is the power factor of the dumped load L; r is the resistance of the power line; $X_L$ is the inductive reactance power line; g is the conductance of the power line; C is capacitance the power line; L is the length of the feeder, f is the system frequency. The selected inputs are the typical data of a 23kV medium voltage feeder of Milano.

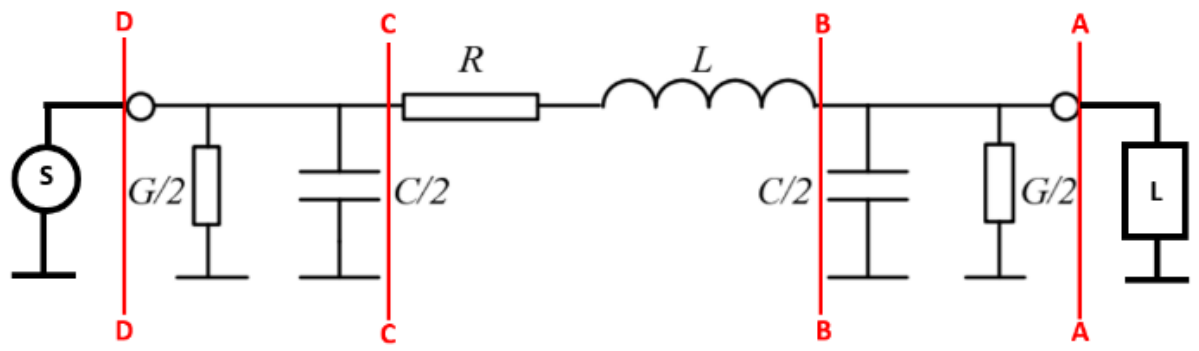

Fig. 17 PI equivalent model of distribution lines.

**Section A_A**

$Q_L = P_L \tan\varphi = 0.66 \text{ MVar}$
$P_{A_A} = P_L = 2\ MW$
$Q_{A_A} = Q_L = 0.66 \text{ MVar}$
$V_{A_A} = V_L = 23 \text{ kV}$
$S_{A_A} = \sqrt{P_{A_A}{}^2 + Q_{A_A}{}^2} = 2.11 \text{ MVA}$
$I_{A_A} = \frac{S_{A_A}}{V_{A_A}} = 91.53 \text{ A}$

**Section B_B**

$V_{B_B} = V_{A_A} = 23 \text{ kV}$
$P_G = g\,L\,V_{B_B}{}^2 = 0$
$$Q_C = -\left(\frac{V_{B_B}{}^2}{\frac{1}{2\,\pi\,f\,\frac{c\,L}{2}}}\right) = -0.187$$
$P_{B_B} = P_{B_B} + P_G = 2\ MW$
$Q_{B_B} = Q_{B_B} + Q_C = 0.47 \text{ MVar}$
$S_{B_B} = \sqrt{P_{A_A}{}^2 + Q_{A_A}{}^2} = 2.05 \text{ MVA}$
$I_{B_B} = \frac{S_{B_B}}{V_{B_B}} = 89.33 \text{ A}$

**Section C_C**

$I_{C_C} = I_{B_B} = 89.33 \text{ kV}$
$P_R = r\,L\,I_{C_C}{}^2 = 0.007$
$Q_L = x_L L\,I_{C_C}{}^2 = 0.005$
$P_{C_C} = P_{B_B} + P_R = 2.01\ MW$
$Q_{C_C} = Q_{B_B} + Q_L = 0.47 \text{ MVar}$
$S_{C_C} = \sqrt{P_{B_B}{}^2 + Q_{B_B}{}^2} = 2.06 \text{ MVA}$
$V_{C_C} = \frac{S_{C_C}}{V_{C_C}} = 23.08 \text{ V}$

**Section D_D**

$V_{D_D} = V_{B_B} = 23.08 \text{ kV}$
$P_G = g\,L\,V_{B_B}{}^2 = 0$
$$Q_C = -\left(\frac{V_{D_D}{}^2}{\frac{1}{2\,\pi\,f\,\frac{c\,L}{2}}}\right) = -0.188$$
$P_{D_D} = P_{C_C} + P_G = 2.01\ MW$
$Q_{D_D} = Q_{C_C} + Q_C = 0.29 \text{ MVar}$
$S_{D_D} = \sqrt{P_{D_D}{}^2 + Q_{D_D}{}^2} = 2.03 \text{ MVA}$

$$I_{D_D} = \frac{S_{D_D}}{V_{D_D}} = 87.81 \text{ A}$$

As shown in Table III, the active and the reactive powers at the point of common coupling with the transmission network ($P_{D_D}$ and $Q_{D_D}$) are both positive. That is not the case when the power demand $P_L$ decreases, for instance, from 2 MW to 1 MW. The reactive power produced by the cables overcompensates the load's reactive demand, which generates a reactive reverse power flow from the distribution to the transmission network.

TABLE III. POWER FLOW, VOLTAGE, AND CURRENT AT THE POINT OF COMMON COUPLING WITH THE TRANSMISSION NETWORK

| $P_L$ [MW] | $\cos\varphi_L$ | $P_{D_D}$ [MW] | $Q_{D_D}$ [MVar] | $V_{D_D}$ [kV] | $I_{D_D}$ [A] |
|---|---|---|---|---|---|
| 2 | 0.95 | 2.01 | 0.29 | 23.08 | 87.81 |
| 1 | 0.95 | 1 | -0.04 | 23.04 | 43.52 |

## REFERENCES

[1] A. Navon, R. Machlev, D. Carmon, A. E. Onile, J. Belikov, and Y. Levron, “Effects of the COVID-19 Pandemic on Energy Systems and Electric Power Grids—A Review of the Challenges Ahead,” *Energies 2021, Vol. 14, Page 1056*, vol. 14, no. 4, p. 1056, Feb. 2021.

[2] C. Bertram *et al.*, “COVID-19-induced low power demand and market forces starkly reduce CO2 emissions,” *Nat. Clim. Chang. 2021 113*, vol. 11, no. 3, pp. 193–196, Feb. 2021.

[3] R. Madurai Elavarasan *et al.*, “COVID-19: Impact analysis and recommendations for power sector operation,” *Appl. Energy*, vol. 279, p. 115739, Dec. 2020.

[4] E. Kawka and K. Cetin, “Impacts of COVID-19 on residential building energy use and performance,” *Build. Environ.*, vol. 205, p. 108200, Nov. 2021.

[5] A. Tingting Xu, B. Weijun Gao, C. Yanxue Li, and D. Fanyue Qian, “Impact of the COVID-19 pandemic on the reduction of electricity demand and the integration of renewable energy into the power grid,” *J. Renew. Sustain. Energy*, vol. 13, no. 2, p. 026304, Apr. 2021.

[6] A. Abulibdeh, “Modeling electricity consumption patterns during the COVID-19 pandemic across six socioeconomic sectors in the State of Qatar,” *Energy Strateg. Rev.*, vol. 38, p. 100733, Nov. 2021.

[7] Q. Wang, S. Li, and F. Jiang, “Uncovering the impact of the COVID-19 pandemic on energy consumption: New insight from difference between pandemic-free scenario and actual electricity consumption in China,” *J. Clean. Prod.*, vol. 313, p. 127897, Sep. 2021.

[8] E. Bompard *et al.*, “The Immediate Impacts of COVID-19 on European Electricity Systems: A First Assessment and Lessons Learned,” *Energies 2021, Vol. 14, Page 96*, vol. 14, no. 1, p. 96, Dec. 2020.

[9] A. Bahmanyar, A. Estebsari, and D. Ernst, “The impact of different COVID-19 containment measures on electricity consumption in Europe,” *Energy Res. Soc. Sci.*, vol. 68, Jul. 2020.

[10] D. Kirli, M. Parzen, and A. Kiprakis, “Impact of the COVID-19 Lockdown on the Electricity System of Great Britain: A Study on Energy Demand, Generation, Pricing and Grid Stability,” *Energies 2021, Vol. 14, Page 635*, vol. 14, no. 3, p. 635, Jan. 2021.

[11] S. Halbrügge, P. Schott, M. Weibelzahl, H. U. Buhl, G. Fridgen, and M. Schöpf, “How did the German and other European electricity systems react to the COVID-19 pandemic?,” *Appl. Energy*, vol. 285, p. 116370, Mar. 2021.

[12] P. M. R. Bento, S. J. P. S. Mariano, M. R. A. Calado, and J. A. N. Pombo, “Impacts of the COVID-19 pandemic on electric energy load and pricing in the Iberian electricity market,” *Energy Reports*, vol. 7, pp. 4833–4849, Nov. 2021.

[13] F. Carere, T. Bragatto, and F. Santori, “A Distribution Network during the 2020 COVID-19 Pandemic,” *12th AEIT Int. Annu. Conf. AEIT 2020*, Sep. 2020.

[14] S. M. Mahfuz Alam and M. H. Ali, “Analysis of COVID-19 Effect on Residential Loads and Distribution Transformers,” *Int. J. Electr. Power Energy Syst.*, vol. 129, pp. 106832–106832, Jul. 2021.

[15] M. Graff and S. Carley, “COVID-19 assistance needs to target energy insecurity,” *Nat. Energy 2020 55*, vol. 5, no. 5, pp. 352–354, May 2020.

[16] Decreto del Presidente del Consiglio dei Ministri 1 marzo 2020, “Ulteriori disposizioni attuative del decreto-legge 23 febbraio 2020, n. 6, recante misure urgenti in materia di contenimento e gestione dell’emergenza epidemiologica da COVID-19.”

[17] Decreto del Presidente del Consiglio dei Ministri 9 marzo 2020, “Ulteriori disposizioni attuative del decreto-legge 23 febbraio 2020, n. 6, recante misure urgenti in materia di contenimento e gestione dell’emergenza epidemiologica da COVID-19, applicabili sull’intero territorio nazionale.”

[18] Decreto del Presidente del Consiglio dei Ministri 13 ottobre 2020, “Ulteriori disposizioni attuative del decreto-legge 25 marzo 2020, n. 19, convertito, con modificazioni, dalla legge 25 maggio 2020, n. 35, recante ‘Misure urgenti per fronteggiare l’emergenza epidemiologica da COVID-19’, e del decreto-legge 16 maggio 2020.”

[19] Decreto del Presidente del Consiglio dei Ministri 18 ottobre 2020, “Ulteriori disposizioni attuative del decreto-legge 25 marzo 2020, n. 19, convertito, con modificazioni, dalla legge 25 maggio 2020, n. 35, recante ‘Misure urgenti per fronteggiare l’emergenza epidemiologica da COVID-19’, e del decreto-legge 16 maggio 2020.”

[20] Decreto del Presidente del Consiglio dei Ministri 25 ottobre 2020, “Ulteriori disposizioni attuative del decreto-legge 25 marzo 2020, n. 19, convertito, con modificazioni, dalla legge 25 maggio 2020, n. 35, recante ‘Misure urgenti per fronteggiare l’emergenza epidemiologica da COVID-19’, e del decreto-legge 16 maggio 2020.”

[21] Decreto del Presidente del Consiglio dei Ministri 4 novembre 2020, “Ulteriori disposizioni attuative del decreto-legge 25 marzo 2020, n. 19, convertito, con modificazioni, dalla legge 25 maggio 2020, n. 35, recante ‘Misure urgenti per fronteggiare l’emergenza epidemiologica da COVID-19’, e del decreto-legge 16 maggio 2020.”

[22] UNARETI S.p.a., “Piano di Sviluppo e Incremento resilienza,” 2021.

[23] IEC 61968, “Application integration at electric utilities - System interfaces for distribution management - Part 11: Common Information Model (CIM),” 2013.

[24] IEC 61970, “Energy Management System application program interface (EMS-aPI) - Part 301: Common Information Model (CIM) Base,” 2013.

[25] ARPA, “Agenzia Regionale per la Protezione dell’Ambiente della Lombardia.” [Online]. Available: https://www.arpalombardia.it/Pages/Meteorologia/Richiesta-dati-misurati.aspx. [Accessed: 29-Nov-2021].

[26] E. Bionda, A. Maldarella, F. Soldan, G. Paludetto, and F. Belloni, “Covid-19 and electricity demand: Focus on Milan and Brescia distribution grids,” *12th AEIT Int. Annu. Conf. AEIT 2020*, Sep. 2020.

[27] G. Iannarelli, A. Bosisio, B. Greco, C. Moscatiello, and C. Boccaletti, “Resilience of the Milan distribution network in presence of extreme events: Covid-19,” in *2020 IEEE International Smart Cities Conference, ISC2 2020*, 2020.

[28] Gestore Mercati Energetici S.p.A., “Relazione annuale 2020,” 2020.

[29] ARERA, “Relazione tecnica valorizzazione transitoria degli sbilanciamenti effettivi in presenza dell’emergenza epidemiologica da COVID-19,” 2020.

[30] ARERA, “Delibera 07 aprile 2020 121/2020/R/eel,” 2020.

[31] ARERA, “Delibera 27 dicembre 2019 568/2019/R/eel,” 2019.

[32] S. Eichner, M. A. Arshad, and R. Grab, “Investigation of network protection coordination with changed reactive power flows in the future grid,” in *NEIS 2020; Conference on Sustainable Energy Supply and Energy Storage Systems*, 2020, pp. 1–6.

[33] ARERA, “Delibera 22 dicembre 2015 646/2015/R/eel,” *2015*. [Online]. Available: https://www.arera.it/it/docs/15/646-15.htm. [Accessed: 20-Feb-2019].

[34] L. Bellani *et al.*, “A supervised classification method based on logistic regression with elastic-net penalization for heat waves identification to enhance resilience planning in electrical power distribution grids,” in *ESREL 2020 PSAM 15*, 2020.

[35] E. Fumagalli, L. Lo Schiavo, S. Salvati, and P. Secchi, “Statistical identification of major event days: An application to continuity of supply regulation in Italy,” *IEEE Trans. Power Deliv.*, vol. 21, no. 2, pp. 761–767, 2006.